\documentclass{IEEEoj}

\usepackage{cite}
\usepackage{amsmath,amssymb,amsfonts}
\usepackage{algorithmic}
\usepackage{graphicx}
\usepackage{textcomp}
\usepackage{booktabs}
\usepackage{setspace}
\usepackage{hyperref}
\usepackage[caption=false,font=footnotesize]{subfig}

\usepackage{xcolor}
\usepackage{orcidlink} 
\AtBeginDocument{\definecolor{ojcolor}{cmyk}{0.93,0.59,0.15,0.02}}
\def\OJlogo{\vspace{-4pt}\hskip-4pt\includegraphics[height=18pt]{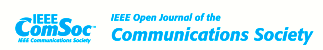}}

\def\BibTeX{{\rm B\kern-.05em{\sc i\kern-.025em b}\kern-.08em
    T\kern-.1667em\lower.7ex\hbox{E}\kern-.125emX}}

\newcommand{\chrome}[1]{{COHERE}}
\newcommand{\TitleStyle}[1]{{\color{ojcolor}#1}}             
\newcommand{\NameStyle}[1]{\textbf{\MakeUppercase{#1}}}      

\begin{document}

\receiveddate{XX Month, XXXX}
\reviseddate{XX Month, XXXX}
\accepteddate{XX Month, XXXX}
\publisheddate{XX Month, XXXX}
\currentdate{23 October, 2025}

\title{\TitleStyle{\chrome{} - Congestion-aware Offloading and Handover via Empirical RAT Evaluation for Multi-RAT Networks}}


\author{%
\NameStyle{Pavan K. Mangipudi}\orcidlink{0000-0002-2116-8174}\IEEEmembership{(Student Member, IEEE)},\;
\NameStyle{Sharon Boamah}\orcidlink{0000-0001-9848-3941}\IEEEmembership{(Student Member, IEEE)},\;
\NameStyle{Lorenz Carvajal},\;
\NameStyle{and Janise McNair}\orcidlink{0000-0002-0448-8723}\IEEEmembership{(Senior Member, IEEE)}
}

\affil{Department of Electrical and Computer Engineering, University of Florida, Gainesville, FL 32611, USA}

\corresp{CORRESPONDING AUTHOR: P. K. MANGIPUDI (e-mail: \href{mailto:pavan.mangipudi@ufl.edu}{pavan.mangipudi@ufl.edu}).}

\authornote{This work was supported by National
Science Foundation under Grant Number 2030122. 
Additional e-mail addresses:
\href{mailto:sharonboamah@ufl.edu}{sharonboamah@ufl.edu},
\href{mailto:lorenzcarvajal@ufl.edu}{lorenzcarvajal@ufl.edu},
\href{mailto:mcnair@ece.ufl.edu}{mcnair@ece.ufl.edu}.}

\markboth{\chrome{} - Congestion-aware Offloading and Handover via Empirical RAT Evaluation for Multi-RAT Networks}{Mangipudi \textit{et al.}}

\begin{abstract}

The evolution of wireless networks and radio access technologies (RATs) has transformed communication from user-driven traffic into a dynamic ecosystem of autonomous systems, including IoT devices, edge nodes, autonomous vehicles, AR/XR clients, and AI-powered agents. These systems exhibit diverse traffic patterns, latency requirements, and mobility behaviors, increasingly operating across overlapping heterogeneous RATs such as 5G, WiFi, satellite, NB-IoT, LoRaWAN, Zigbee, etc. This multi-RAT coexistence creates opportunities for intelligent access, mobility, and routing strategies. However, most mobility decisions still rely heavily on RSSI, which neglects RAT-specific features, congestion, queuing delays, and application needs, favoring high-power links over optimal ones. To address this gap, we propose \ chrome{}  (Congestion-aware Offloading and Handover via Empirical RAT Evaluation), a multi criteria framework for dense multi-RAT networks. \ chrome{}  enhances RSSI with multiple criteria and applies the Technique for Order of Preference by Similarity to the Ideal Solution (TOPSIS) to rank available RATs. Criteria weights are determined using both subjective (operator-driven) and objective (measurement-based) approaches. Based on this ranking, \ chrome{}  performs intelligent cross-RAT offloading to reduce congestion on over-utilized links. We evaluate \ chrome{}  in a dense SDN-controlled 5G/WiFi Multi-RAT environment using Mininet WiFi. Compared to RSSI-only handover, \chrome{} reduces the load on the congested RAT by up to 32\%, reduces total handovers by 25\%, lowers handovers to the congested RAT by 55\%, and improves link delay by up to 166\%, while maintaining comparable or up to 11\% higher throughput. These results demonstrate that guarded, multi-criteria decision-making can exploit RAT coexistence to deliver robust, congestion-aware performance across heterogeneous deployments.


\end{abstract}


\maketitle

\section{Introduction}

The Sixth generation (6G) of cellular networks is anticipated to provide enhanced spectral efficiency, energy efficiency, native AI integration, and and further improvements in latency and data rates compared to previous generations. 6G also aims to extend connectivity among humans, machines, and hybrid systems through emerging technologies such as the Internet of Things (IoT), vehicular networks, drones, satellites, and other heterogeneous Radio Access Technologies (RATs).Moreover, there is an increasing densification of the cellular Radio Access Network (RAN) to support this heterogeneous wireless access environment. Diverse deployment scenarios involving macro cells, small cells, indoor coverage solutions, and private networks will enable service providers to expand coverage and enhance connectivity~\cite{zaididense5G2020}. This coexistence of dense and diverse multi-RAT environments has created an ecosystem that can be leveraged to address 6G connectivity challenges through coordinated and efficient multi-RAT mobility management.

However, current multi-RAT handover mechanisms remain predominantly based on signal strength indicators (e.g., RSRP/RSRQ/SINR in NR and RSSI/SNR in WiFi). Existing approaches often overlook the distinctive merits and operational characteristics of heterogeneous RATs, as well as the variations in their parameters and performance capabilities. Each RAT is engineered for specific purposes and operational requirements, which shape its spectrum usage, PHY/MAC design, and deployment assumptions. Consequently, RATs exhibit distinct propagation behaviors, channel constraints, and performance profiles, rendering RSSI alone inadequate for reliable handover decisions. Thus, a handover mechanism that relies solely on signal strength parameters, without incorporating performance indicators relevant to Quality of Service (QoS) and Quality of Experience (QoE), fails to capture the unique strengths and limitations of each RAT, leading to suboptimal handover performance.

From an implementation standpoint, heterogeneous RATs can be integrated through both standardized and research/implementation methods. These mechanisms chiefly solve interworking and control/data-plane integration rather than the handover decision itself. Examples span standardized mechanisms—such as LTE–WiFi Aggregation (LWA)~\cite{ltewifiaggregation3gpp},\cite{ltewifiaggregation} and Dual Connectivity (DC) in 5G NR\cite{5gnrDualConn3gpp}—and, in 5G, WiFi/NR interworking via N3IWF together with ATSSS. Research and implementation efforts also explore SDN-based coordination to unify control across RATs~\cite{KHATURIA2021108412,liyanage2017sdnoffloading,wang2016convergence}. These approaches primarily provide interworking and an integrated control-plane architecture, but they do not prescribe a unified, RAT-aware, multi-criteria handover policy. In practice, mobility often defaults to signal-strength–driven triggers within each RAT or to core-level policies with limited RAN visibility, leaving cross-RAT handovers largely RAT-agnostic~\cite{KHATURIA2021108412}.


In a preliminary study \cite{mangipudi2023LANMAN}, we implemented a proof-of-concept MCDM handover using entropy-weighted TOPSIS with only RSSI and link delay as criteria and a threshold-checked “stand-in” guard to avoid pathological RAT specific selections. This resulted in reduced handover failure ratio, improved throughput, and significantly lowered delay compared to an RSSI-based baseline. However, it did not consider load on each multi-RAT node or quantify the impact of offloading through direct evaluation of access-node load after the handover or offloading. Furthermore, it relied solely on entropy-based weighting, which can be brittle under extreme operating conditions, and was evaluated at modest scale ($\leq 16$ UEs) in emulation. Building on that foundation, this work provides an offloading centric framework by adding access-node load as an optimization criterion and directly measuring access-node load redistribution to assess offloading efficacy. Additionally, this work also compares subjective and objective weighting by integrating AHP-TOPSIS alongside Entropy-TOPSIS within a single unified ranking pipeline, and scales evaluation to dense scenarios (up to 64 UEs) while retaining the RAT-aware offloading guard. 

Accordingly, we propose \chrome{}, Congestion-aware Offloading and Handover via Empirical RAT Evaluation and make the following contributions:

\begin{itemize}
    \item 
    A multi-criteria handover and offloading approach that considers link delay, signal strength (RSSI), and load on the access node, explicitly accounting for inherent asymmetry among RATs (e.g., higher transmit power and potential congestion on macro links versus lower latency and and lower power in local access technologies).
    \item A generalized Multi-Criteria Decision-Making (MCDM) based handover and offloading framework for heterogeneous multi-RAT networks, that addresses the limitations of signal-strength–centric handovers in scenarios where different RATs exhibit diverse performance characteristics, and is deployable atop both standards-based (e.g., N3IWF+ATSSS) and SDN-based integrations.

    \item Integration of two weighing strategies into the Technique for Order of Preference by Similarity to the Ideal Solution (TOPSIS) decision making framework, facilitating offloading in extreme network scenarios. Analytical Hierarchical Process (AHP) based weighting for environments with unpredictable network behavior, allowing operator preferences to guide decision making. Entropy-based weighting for structured or predictable environments, where data-driven variability is used to prioritize the most informative criteria. Both operate under a unified MCDM framework, eliminating the need for algorithm switching at runtime.
    \item A dynamic offloading mechanism that removes strongest-signal bias by selecting an alternative qualifying nodes (e.g., nearby WiFi APs) that satisfies feasibility constraints (e.g., a minimum signal-strength threshold). This facilitates avoiding the congested high-power RATs and improves multi-RAT handovers in congested, load-imbalanced scenarios.
\end{itemize}



To the best of our knowledge, few implementations apply MCDM-based ranking to handovers in heterogeneous multi-RAT networks. Furthermore, this is the first approach to use a robust MCDM-based handover to remove strongest-signal (high RSSI) bias, enabling effective multi-RAT offloading.

The rest of the paper is organized as follows. Section~\ref{sec_bg} provides related work on SDN-based multi-RAT networks, corresponding handover methods, and the limitations associated with each approach. Section~\ref{sec_architecture} covers the system description of the Multi RAT network and Section~\ref{sec_mcdmusecases} provides a description of how MCDM based handover works, while identifying limitations specifically for offloading. Section~\ref{sec_ho_algo} gives a comprehensive description of the proposed \chrome{} framework. Section~\ref{sec_perfAnalysis_sim} provides the details on the simulation of the multi-RAT network and performance evaluation of the proposed \chrome{} framework, using Mininet WiFi\cite{mininetwifi}. Finally, Section~\ref{sec_conclusion} concludes the paper and provides directions for future research.

\section{Related Work} 
\label{sec_bg}

Multi-RAT networks have received significant attention in recent years, with various implementations enabling coexistence between heterogeneous access technologies such as LTE, 5G NR, WiFi, and satellite networks. However, most of these efforts focus on enabling connectivity and interworking between RATs, rather than on the mobility or handover decision itself.

\subsection{Multi-RAT network architectures}
A number of works focus on architectural solutions that facilitate multi-RAT integration through SDN-based or standards-based mechanisms, but do not prescribe how mobility decisions should be made across those RATs. For example, LTE–WiFi Aggregation (LWA), Dual Connectivity (DC), and 5G interworking via N3IWF and ATSSS allow non-3GPP RATs to interoperate with 3GPP core networks~\cite{ltewifiaggregation3gpp, ltewifiaggregation, 5gnrDualConn3gpp}. These approaches emphasize control and data-plane integration, supporting path continuity, session management, and user-plane routing across RATs. Similarly, SDN-based approaches aim to unify the control plane and manage flow-based routing between different RATs~\cite{KHATURIA2021108412, liyanage2017sdnoffloading, wang2016convergence}. While these architectures provide the foundation for seamless multi-RAT connectivity, they generally do not define how handover decisions are made or optimized across RATs. In most cases, mobility still relies on per-RAT signal strength–based triggers or static rules applied at the core level, with limited visibility into the access network or performance differences between RATs~\cite{KHATURIA2021108412}.

\subsection{Multi-RAT systems with single decision criteria}
Some works claim to support multi-RAT handover or offloading, but the underlying selection mechanisms either rely on a single performance metric—typically signal strength—or fail to define a decision policy altogether. For instance, the hierarchical SDN-based handover architecture proposed in~\cite{Alfoudi2019seamless} facilitates seamless transitions between WiFi, LTE, and 5G, and reports gains in signaling overhead and handover delay. However, it does not detail the criteria used to select the target RAT or access point, and as a result, the decision process defaults to signal strength based metrics. A similar issue arises in~\cite{liyanage2017sdnoffloading}, which describes an operator-assisted offloading framework between LTE and WiFi using separate SDN controllers. While the architecture enables multi-RAT offloading, it does not specify how compatible networks are identified or ranked, nor how the controllers use performance measurements to guide decisions. Another example is~\cite{wang2016convergence}, which introduces "virtual middle-boxes" to maintain IP continuity during vertical handovers. While this ensures service stability, it avoids the more complex issue of performance-based target selection during handover. In all these cases, multi-RAT integration is achieved, but the actual decision logic is either limited to a single metric or is unspecified, and therefore fails to reflect the diverse characteristics and performance profiles of different RATs. This highlights a key limitation in the literature, while multi-RAT mobility is enabled, the handover decisions are often reduced to RSSI-based logic, missing the potential for more nuanced, RAT-aware, multi-criteria optimization.

\subsection{Multi-criteria based handover in singular RATs}
A separate class of work focuses on improving handover decisions using multiple performance criteria, but does so within the context of a single RAT. These studies are important as they demonstrate the utility of multi-criteria decision-making (MCDM) frameworks, yet they do not extend this logic to heterogeneous RAT environments. For example, the work in~\cite{ciciouglu2021multi} applies an entropy-weighted Simple Additive Weighting (SAW) method for handover in ultra-dense 5G small cells, considering metrics such as SINR, bandwidth, and user density. This approach reflects dynamic network conditions and adapts well to small cell environments, but it is designed solely for intra-RAT mobility within 5G networks and does not consider the asymmetries between different technologies such as WiFi and NR. Similarly, other works such as~\cite{prados2016handover, rizkallah2018sdnverticalho, abdulghaffar2021modeling} explore improvements to handover control or routing policies in LTE or NR, sometimes acknowledging non-RSSI criteria, but do not implement or evaluate their selection mechanisms in a multi-RAT context. In these cases, the assumption is often a homogeneous network with uniform propagation characteristics and performance expectations, which limits their applicability in heterogeneous deployments. While these contributions validate the value of using multiple criteria to guide mobility decisions, they do not address the specific challenges introduced by RAT diversity—such as varying link delays, transmit power, or congestion behavior—nor do they offer a strategy for offloading between dissimilar technologies. As such, they fall short of addressing the central challenge of RAT-aware handover in multi-RAT networks.
Other works do explore handover decision-making using criteria beyond signal strength, particularly in dense or heterogeneous environments. In~\cite{ciciouglu2021multi}, the authors apply a multi-criteria decision-making (MCDM) method using entropy-weighted Simple Additive Weighting (SAW) to improve handover selection in ultra-dense 5G small cell deployments. The approach considers parameters like SINR, bandwidth, and user density. However, it is limited to a single RAT (5G) and does not address the unique challenges posed by multi-RAT coexistence, such as varying delay profiles and access-node capabilities. Similarly, while~\cite{rizkallah2018sdnverticalho} discusses vertical handovers in cellular networks, the approach remains largely theoretical and does not provide a concrete, RAT-aware selection mechanism.

In our prior work~\cite{mangipudi2023LANMAN}, we we partially address this gap by implementing a proof-of-concept entropy-weighted TOPSIS handover across 5G and WiFi using RSSI and link delay. This also included an offloading algorithm that selects WiFi nodes to reduce the usage of 5G spectrum and a threshold-checked feasibility guard to avoid pathological selections. However, that study did not incorporate access-node load into the decision process, nor did it quantify offloading effects via direct measurements of load redistribution across access nodes or the number of handovers relative to an RSSI-based baseline. It relied solely on entropy-based weighting, which can become brittle when one criterion exhibits disproportionately high variance, and the evaluation was limited to modest-scale emulation. Importantly, although a rudimentary guard was present, the prior work did not isolate or demonstrate the failure mode of an MCDM-only policy (i.e., without RAT-specific feasibility constraints) under congested deployments; as a result, the necessity and design of a RAT-aware guard were not established empirically.

In this work, we address these gaps by considering access-node load alongside RSSI and link delay to enable direct assessment of offloading efficacy. We also introduce subjective (AHP) and objective (entropy) weighting within a single TOPSIS pipeline to improve robustness across operating regimes; and formalizing a RAT-specific feasibility guard that mitigates strongest-signal bias. We explicitly analyze why an MCDM-only approach can perversely favor high-power, high-load macro links in congested UE distributions—thereby failing to relieve RAN congestion—and how the RAT specific offloading guard can help fix this. We evaluate the unified framework under both normal and congested UE placements in Mininet-WiFi, scaling to dense scenarios (up to 64 UEs) and reporting its impact on delay, throughput, access-node load redistribution, and handover dynamics.

\section{Background}

\subsection{Multi Criteria Decision Making (MCDM)} 
\label{sec_bg_mcdm}

Multi-Criteria Decision Making (MCDM), as the name implies, is a decision-making framework used to select the most suitable alternative from a set of candidates by ranking based on multiple evaluation criteria. The simplest MCDM method is a weighted sum, wherein the criterion weights, normalized to sum to one, are used to compute a weighted score for each alternative. This is done by multiplying normalized criterion values from each alternative with the corresponding criterion weight, and adding the results to generate a weighted sum for each alternative or candidate. The alternative with the highest weighted sum is selected as the final choice. 

The MCDM process can be broken down into two steps: (1) obtaining criterion weights and (2) decision-making \cite{mcdm_in_networks,mcdmbook}. Two widely used and contrasting weighing approaches are subjective or (fixed) weighting and objective (dynamic) weighting. Subjective weights are determined before the decision-making process, based on operator preference or pre-defined heuristics that compare the relative importance of each criterion. They do not change throughout the process. In contrast, objective weights are calculated dynamically, at each decision-making instance, based on the criterion values obtained from each alternative during the given decision-making instance. When the criterion values of each alternative change, the weights calculated change according to the objective heuristic defined. The subjective weighing technique used in this work is the Analytical Hierarchical Process (AHP) based weighing, and the objective weighing approach is the entropy-based weighing approach. These methods are particularly well-suited to the complexities of multi-RAT environments. AHP is chosen for its ability to capture criteria priority through structured pairwise comparisons, allowing nuanced differentiation between criteria like RSSI, delay, and load based on RAT-specific behaviors. Entropy is selected for its data-driven nature, assigning weights objectively based on the variability of observed measurements, making it well-suited for dynamic network environments.

\subsubsection{AHP based weighting} \label{sec_bg_mcdm_ahp} 
For AHP, the next step involves constructing a pairwise comparison matrix $\textbf{P} = p_{ij}$ for the M criteria. This matrix is of size $MXM$, where each element $p_{ij}$ represents the relative importance of criterion $i$ compared to criterion $i$, as determined by expert judgment. The diagonal elements of the matrix are always 1, as each criterion is equally important to itself, and the matrix satisfies the reciprocal property: $p_{ij} = 1/p_{ji}$ entities. To guide the assignment of relative importance values, the Saaty scale is used \cite{ahp}. 

Once the matrix $\textbf{P}$ is constructed, the weights are computed by normalizing the matrix column-wise and then averaging across each row. Specifically, each element is divided by the sum of its column:

\begin{equation} \label{eq:ahp_pairwise_norm}
n_{ij} = \frac{p_{ij}}{\sum_{i=1}^{M}p_{ij}}
\end{equation}
Then, the weight for each criterion $i$ is given by:
\begin{equation} \label{eq:ahp_weight}
w_{i} = \frac{1}{M}{\sum_{j=1}^{M}n_{ij}}
\end{equation}

This produces the subjective weight vector for AHP $\textbf{w}_{AHP} = [w_{1},w_{2}, ... w_{M}]^T $. A consistency check is then performed to ensure that the judgments are logically coherent.

\subsubsection{Entropy based weighting} \label{sec_bg_mcdm_entropy}

Entropy-based weighting begins with the construction of the decision matrix $\textbf{X}$ from the available alternatives (or candidate nodes). Each element $x_{ij}$ represents the value of the $j^{th}$ criterion for the $i^{th}$ alternative, and N is the number of alternatives, and M is the number of criteria. Once the decision matrix $\textbf{X}$ is formed, it is normalized—using linear normalization to create a normalized decision matrix $\textbf{R} = r_{ij}$. 

\begin{equation} \label{eq:entropy_linear3_normalization}
r_{ij} = \frac{x_{ij}}{\sum_{i=1}^{n} x_{ij}}
\end{equation}

Next, the entropy for each criterion \( j \) is computed as:

\begin{equation} \label{eq:entropy}
e_j = -k \sum_{i=1}^{n} r_{ij} \cdot \ln(r_{ij}), \quad \text{where } k = \frac{1}{\ln(n)}
\end{equation}

If \( r_{ij} = 0 \), the corresponding term is taken as zero to avoid undefined logarithmic operations.

Next, the degree of diversification \( d_j \) is calculated as follows:

\begin{equation} \label{eq:diversification}
d_j = 1 - e_j
\end{equation}

Finally, the normalized weights \( w_j \) for each criterion are computed as:

\begin{equation} \label{eq:ent_weights}
w_j = \frac{d_j}{\sum_{j=1}^{m} d_j}
\end{equation}

The resulting weight vector is \( \mathbf{w}_{Entropy} = [w_1, w_2, \dots, w_m]^T \), where criteria with higher variability (and thus higher discriminatory power) are assigned greater weight.

\subsubsection{Subjective vs. Objective Weighting for Multi-RAT Handover} \label{sec_bg_mcdm_weighting_compare}

Both AHP and entropy-based weighting ultimately assign relative importance to decision criteria; the key difference lies in what drives those weights. AHP is goal- and policy-driven, while entropy is preference-free and data-driven. Using AHP, the intended handover objective and acceptable trade-offs can be explicitly encoded through pairwise comparisons (e.g., RSSI vs. access-node load, access-node load vs. delay), with a consistency check ensuring that these preferences form a coherent policy. This allows the framework to prioritize, for instance, RSSI over load on the access node (or vice versa), even when the instantaneous measurements do not show strong differences across candidates.
In contrast, entropy-based weighting assigns higher importance to criteria that exhibit greater information content or dispersion across candidates at each decision epoch. For example, if RSSI values are relatively constant but instantaneous link delay varies widely, entropy assigns greater weight to delay, emphasizing the most discriminative criterion in that moment. This adaptive behavior can yield QoS gains when dispersion reflects meaningful performance variation. However, entropy cannot directly encode operator intent, and may overweight noisy or volatile criteria, especially when the statistical ranges are uncertain or non-stationary.
For this reason, we treat the two as complementary approaches, each suited to different network behaviors. AHP provides a stable, goal-aligned subjective weighting scheme that is useful in uncertain or policy-sensitive scenarios. Entropy offers a responsive, objective mechanism that adapts to real-time measurements, making it well-suited to dynamic environments with predictable measurement characteristics.

\subsection{TOPSIS} \label{sec_bg_mcdm_topsis} 
The Technique for Order of Preference by Similarity to Ideal Solution (TOPSIS) is a multi-criteria decision-making method that ranks alternatives based on their geometric distance from an ideal solution. In this approach, weights for the criteria are assumed to be precomputed using a separate method such as AHP or entropy-based weighting. The decision matrix \( \mathbf{X} = [x_{ij}] \) is first normalized using vector normalization to construct the normalized decision matrix \( \mathbf{R} = [r_{ij}] \):

\begin{equation} \label{eq:topsis_vector_normalization}
r_{ij} = \frac{x_{ij}}{\sqrt{\sum_{i=1}^{n} x_{ij}^2}}
\end{equation}

Using the precomputed weights \( \mathbf{w} = [w_1, w_2, \dots, w_m]^T \), the weighted normalized matrix \( \mathbf{V} = [v_{ij}] \) is computed as:

\begin{equation} \label{eq:norm_weighted_matrix}
v_{ij} = w_j \cdot r_{ij}
\end{equation}

Next, the ideal solution \( \mathbf{v}^+ \) and the negative-ideal solution \( \mathbf{v}^- \) for benefit and cost criteria \( J_{\text{benefit}}, J_{\text{cost}} \), respectively, are calculated using the following equations:

\begin{equation}
\label{eq:ideal_solutions}
\begin{aligned}
v_j^+ &=
\begin{cases}
\max_i v_{ij}, & \text{if } j \text{ is a benefit criterion},\\
\min_i v_{ij}, & \text{if } j \text{ is a cost criterion},
\end{cases}\\[6pt]
v_j^- &=
\begin{cases}
\min_i v_{ij}, & \text{if } j \text{ is a benefit criterion},\\
\max_i v_{ij}, & \text{if } j \text{ is a cost criterion}.
\end{cases}
\end{aligned}
\end{equation}

Benefit criteria are those to be maximized, while cost criteria refer to those to be minimized. The ideal solution acts as a benchmark for what an optimal solution would look like. Each real alternative is evaluated by its proximity to this ideal point in a geometric sense. Similarly, the negative-ideal solution serves as a contrast point. An alternative that is far from the worst case (negative ideal) is considered more favorable, especially if it’s also close to the ideal. So, the Euclidean distance of each alternative from the ideal and negative ideal solutions is calculated to determine the proximity to the ideal solution and negative ideal solution, respectively, using the following equation:

\begin{equation}
\label{eq:distance_idealsols}
\begin{aligned}
S_i^+ &= \sqrt{ \sum_{j=1}^{m} (v_{ij} - v_j^+)^2 },\\[4pt]
S_i^- &= \sqrt{ \sum_{j=1}^{m} (v_{ij} - v_j^-)^2 }.
\end{aligned}
\end{equation}

The relative closeness coefficient \( C_i \) for each alternative is computed as follows:

\begin{equation} \label{eq:closeness_coeff}
C_i = \frac{S_i^-}{S_i^+ + S_i^-}
\end{equation}

Finally, alternatives are ranked in descending order of their closeness coefficient \( C_i \). A higher \( C_i \) indicates a better alternative.

Our proposed \chrome{} framework uses weights obtained using both AHP and Entropy methods, respectively, with TOPSIS to produce two distinct MCDM-based handover and offloading approaches.

\subsection{Software Defined Networking}

Software-Defined Networking (SDN) offers a flexible and programmable architecture that is particularly well-suited to the integration and management of heterogeneous wireless networks, such as multi-RAT (Radio Access Technology) environments. A defining feature of SDN is its separation of the network’s control logic from the data forwarding functions \cite{wirelesssdnsurvey}. This architectural shift allows centralized control of distributed network elements, which is especially valuable in managing diverse and dynamic RATs.

\begin{figure}[ht!]
\centerline{\includegraphics[width=\columnwidth]{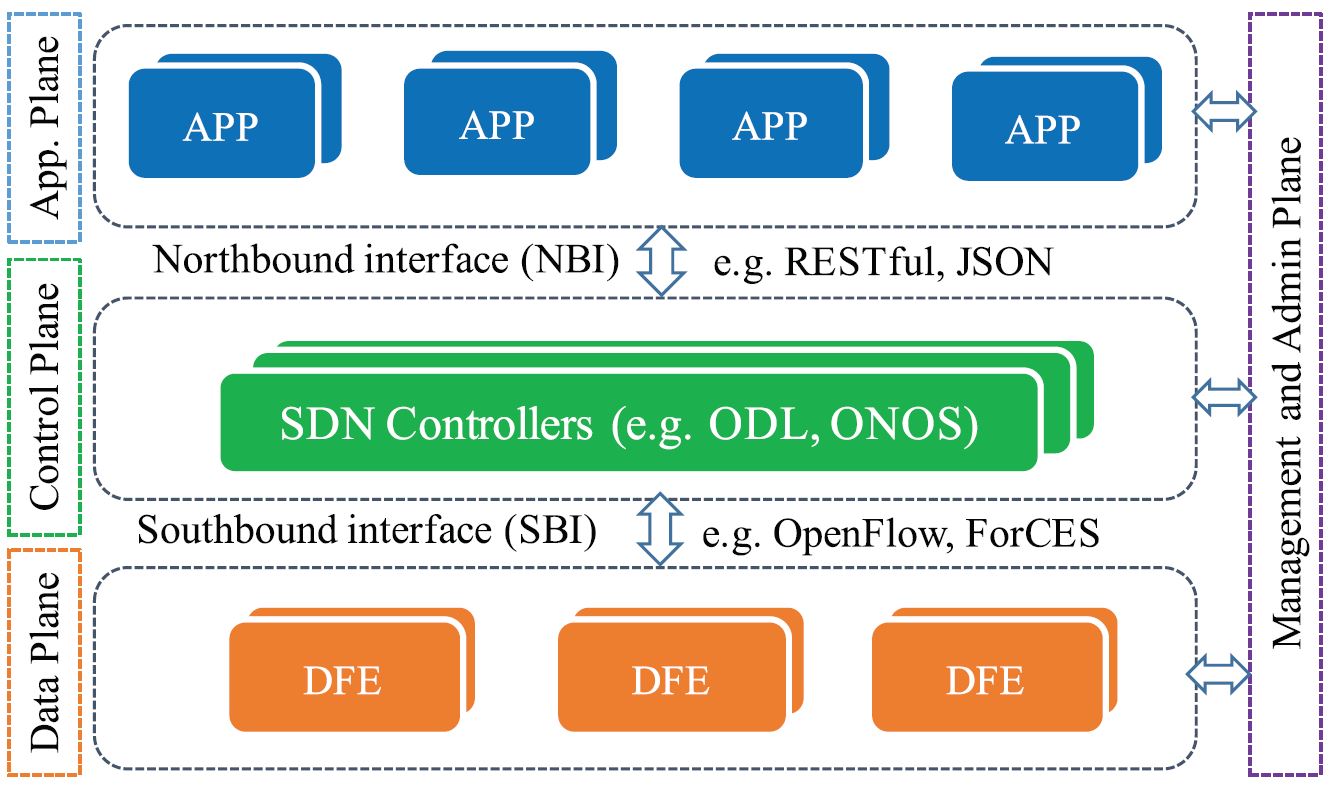}}
\caption{A high level SDN architecture \cite{nguyen2017sdnnfvltesurvey}.}
\label{SDNarch}
\end{figure}

As illustrated in Fig.~\ref{SDNarch}, the SDN architecture comprises multiple logical planes. At the center is the \textit{control plane}, which contains one or more SDN controllers with a global view of the network. These controllers manage forwarding decisions, network policies, and resource coordination across multiple access technologies.

Above the control plane lies the \textit{application plane}, where high-level services such as handover management, routing optimization, and load balancing reside. These applications define the operational policies and pass their requirements to the control plane via northbound interfaces (NBIs).

The \textit{data plane} consists of physical or virtual forwarding devices—such as switches and routers—that execute the control logic received from the SDN controller. These devices are connected to the controller through southbound interfaces, enabling fine-grained control over packet forwarding behavior across heterogeneous links. An additional \textit{management plane} oversees tasks such as monitoring, configuration, and provisioning of network resources.

The first protocol to standardize this interaction, \textit{OpenFlow} \cite{openflow}, remains a key enabler of SDN functionality. Managed by the Open Networking Foundation (ONF) \cite{onf}, OpenFlow allows controllers to install "flows" in forwarding devices. Each flow rule defines matching fields, priority levels, counters, timeouts, and packet-handling instructions. These flow entries are organized in flow tables, and are responsible for handling packet classification and forwarding actions.

Through OpenFlow, the SDN controller can proactively or reactively manage the flow tables of data plane devices. The controller’s centralized and global perspective allows it to make intelligent decisions regarding resource allocation and routing, which is essential in multi-RAT scenarios where seamless coordination and load-aware handovers across access technologies are required.

\section{System Architecture}\label{sec_architecture}

Multi-RAT integration can be realized in at least two practical ways. First, the 3GPP standards path integrates WiFi and 5G RAN through RAT-specific entities. Non-3GPP Interworking Function (N3IWF) for untrusted WiFi networks  and Trusted Non-3GPP Gateway (TNGF) for trusted WiFi networks, with traffic steering handled in the core via ATSSS. This approach is standards compliant and can also host our proposed \chrome{}  framework by mapping the ranking output to RAT selection and traffic steering rules. However, it has some architectural and implementation constraints. Specifically, it requires separate interworking functions for each RAT, increases signaling overhead due to per-RAT registration and tunnel establishment, and limits routing flexibility as decisions are made in the 5G core with limited visibility into RAN-level metrics. In the case of N3IWF, additional encapsulation (GRE/IPSec) is introduced in both control and user planes, further complicating the data path and reducing efficiency~\cite{KHATURIA2021108412}.

Second, an SDN-based realization unifies interconnection at the RAN edge using a centralized SDN controller and an OpenFlow-enabled switch layer. In this design, gNBs and WiFi APs attach to OpenFlow-enabled switches, and the SDNC programs forwarding based on real-time RAN telemetry (e.g., RSSI, link delay, and load on the connected node). This provides per-flow, per-RAT control, supports decoupled uplink/downlink routing and local breakout when needed, and exposes the timely measurements our MCDM handover requires. Similar SDN based implementation of multi-RAT networks is seen in works such as \cite{Alfoudi2019seamless,wang2016convergence,liyanage2017sdnoffloading} to simplify network management and orchestration. Therefore, in this paper we do not extensively cover this aspect and instead concentrate on the handover decision algorithm.

\begin{figure}[ht!]
 \centerline{\includegraphics[width=\columnwidth]{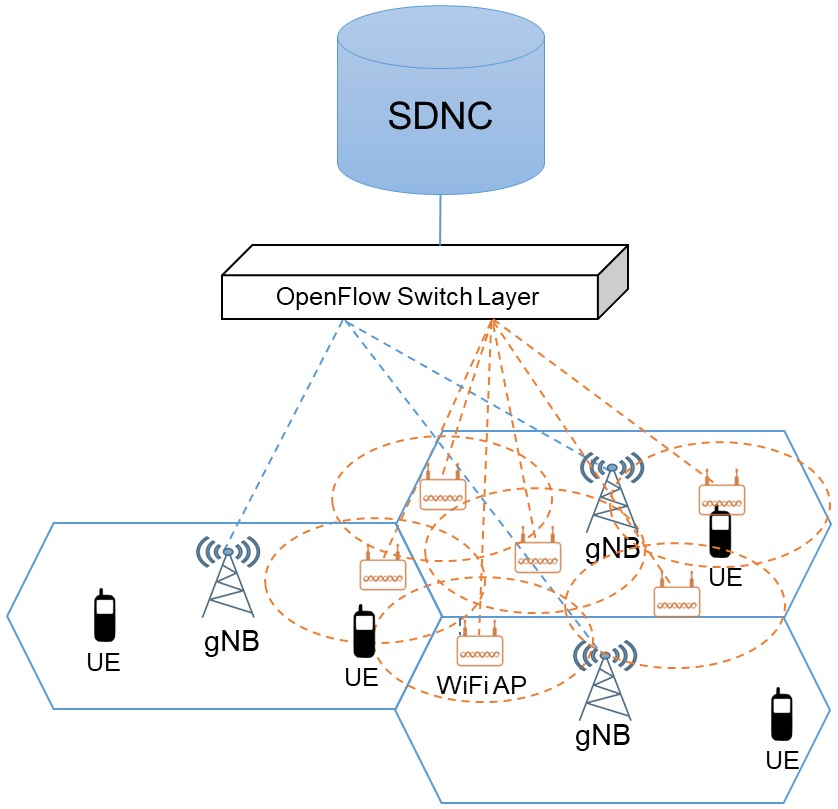}}
\caption{SDN-based Multi RAT network architecture}
\label{architecture}
\end{figure}

Fig. \ref{architecture} shows the architecture of the proposed SDN-based multi-RAT network with WiFi and 5G nodes. A single centralized SDN controller is used to monitor the network and installs forwarding rules on the OpenFlow-enabled switches in the switch layer. The gNBs and the WiFi APs are connected to the OpenFlow-enabled switches, which forward user-plane traffic according to the flow rules. Each switch is considered to be compatible with each of the respective gNBs and APs. The UEs are assumed to be capable of using both WiFi and 5G RATs~\cite{wwangcoexist2023}. The RSSI of each UE with respect to the nodes in its coverage area, the average link delay with respect to each node, and the load on each node are collected periodically to facilitate an MCDM-based handover. While handover optimization and mobility management applications can occur on the SDN controller, the required modification of the OpenFlow messages, e.g., as seen in \cite{ciciouglu2021multi}, is beyond the scope of this work. So, in this work, SDN is used to provide the interconnection between the two RATs at the IP level and to optimize the packet forwarding, as in \cite{prados2016handover, rizkallah2018sdnverticalho}.

\section{MCDM based handover in multi-RAT networks} \label{sec_mcdmusecases}

This section presents a case study of how MCDM based handover algorithms described in Section~\ref{sec_bg_mcdm}, can be used in various multi-RAT scenarios and to identify their limitations. Specifically, we examine the performance of AHP- and Entropy-based TOPSIS schemes, using RSSI, access-node load and link delay as the performance criteria to rank the available nodes in a UE's coverage area. First, we present present two representative use cases from realistic multi-RAT deployments and analyze how each of the two weighting techniques performs, highlighting their respective advantages. Then, we show that relying on a single approach fails to address offloading comprehensively and often requires static reweighting to achieve the desired outcome. This analysis helps to identify the limitations of using these MCDM-based approaches directly in multi-RAT networks.

The multi-RAT architecture described in Section~\ref{sec_architecture} comprises of 5G based macrocells operating on 2.4GHz, and multiple distributed smaller WiFi APs operating on 5GHz. In this work, we characterize the multi-RAT networks using RSSI, access-node load, and instantaneous link delay. So, the 5G gNBs and WiFi APs can be expected to have unique, deterministic characteristics for each of the three criteria. The 5G macrocells provide coverage for a vast geographical area through high transmit powers. So, the UEs closer to them will have the propensity to select the cellular network due to higher RSSI associated with the higher transmit power, regardless of the load on the network. Conversely, the WiFi APs provide smaller coverage areas due to relatively lower transmit powers, while having comparatively lower loads, but are uniformly distributed across the geographical area, which can provide lower delays. For such a network topology, two important use cases can be identified to explain the applicability and limitations of the MCDM-based algorithms in varying network conditions: UEs that are farther from the 5G gNBs and UEs that are closer to the 5G gNBs. For each use case, we describe how the AHP-based MCDM and Entropy-based MCDM scheme work by providing the weights and final rankings, and derive the preferred method for each case.

\subsection{Distributed UEs — RSSI-Prioritized Use Case }\label{sec_perfAnalysis_usecase1} 
When UEs are spatially distributed across the network and a UE lies within the coverage of both the gNB and nearby WiFi APs, their received signal strengths can be comparable. Given the typical spatial density of WiFi, we consider two WiFi APs within the UE’s coverage alongside one gNB. Because the 5G base station serves a larger area, it will generally have more associated stations than any single WiFi AP at a given time. However, depending on link conditions, the WiFi APs may offer stronger RSSI and lower delay than the gNB. Based on these characteristics, we construct an example decision matrix for this case, as shown in Table~\ref{tab:uc1-raw}.

\begin{table}[h!]
\caption{Decision matrix for RSSI use case}
\label{tab:uc1-raw}
\centering
\begin{tabular}{l|ccc}
\toprule
Candidate & RSSI (dB) & Load & Delay (sec) \\
\midrule
gNB & -80 & 9 & 0.035 \\
AP5 & -84 & 0 & 0.032 \\
AP7 & -79 & 2 & 0.023 \\
\bottomrule
\end{tabular}
\end{table}

For these given values of RSSI, Load and Delay, we compute the best alternative using the Entropy based TOPSIS and AHP based TOPSIS methods to demonstrate the performance of each method in the selected use case. 

\paragraph{Entropy based TOPSIS} The entropy-based approach is applied to RSSI use case as follows. It begins with linear normalization of the raw decision matrix, shown in Table~\ref{tab:uc1-raw}, using eqn~\eqref{eq:entropy_linear3_normalization}, to obtain the normalized values, shown in Table~\ref{tab:uc1-norm}.

\begin{table}[h!]
\caption{Normalized decision matrix for RSSI use case (linear normalization)}
\label{tab:uc1-norm}
\centering
\begin{tabular}{l|ccc}
\toprule
Candidate & RSSI & Load & Delay \\
\midrule
gNB & 0.37636 & 0.81818 & 0.38889 \\
AP5 & 0.14983 & 0.00000 & 0.35556 \\
AP7 & 0.47381 & 0.18182 & 0.25556 \\
\bottomrule
\end{tabular}
\end{table}

From the normalized matrix, the entropy values for each criterion are calculated using eqn~\eqref{eq:entropy}. These are used to derive divergence values using eqn~\eqref{tab:uc1-ahp-dist}. Table~\ref{tab:uc1-entropy} shows the respective values for entropy and divergence. Finally, entropy-based weights are obtained using eqn~\eqref{eq:ent_weights}, and are shown in Table~\ref{tab:uc1-entropyweights}. This concludes the entropy based weight calculation. 

\begin{table}[h!]
\caption{Entropy and divergence values for RSSI use case}
\label{tab:uc1-entropy}
\centering
\begin{tabular}{l|cc}
\toprule
Criterion & Entropy \( e_j \) & Divergence \( d_j \)  \\
\midrule
RSSI  & 0.91580 & 0.08420  \\
Load  & 0.43158 & 0.56842  \\
Delay & 0.98635 & 0.01365  \\
\bottomrule
\end{tabular}
\end{table}

\begin{table}[h!]
\caption{Entropy based weights for RSSI use case}
\label{tab:uc1-entropyweights}
\centering
\begin{tabular}{l|c}
\toprule
Criterion & Entropy based weights \( \mathbf{w}_{Entropy} \) \\
\midrule
RSSI & 0.12637 \\
Load & 0.85314 \\
Delay & 0.02048 \\
\bottomrule
\end{tabular}
\end{table}

The next phase is to use the entropy based weights to obtain the TOPSIS based rankings. First, the weighted decision matrix shown in Table~\ref{tab:uc1-weighted} is obtained using eqn~\eqref{eq:norm_weighted_matrix}, by multiplying each row of the normalized decision matrix shown in Table~\ref{tab:uc1-norm} with the calculated entropy based weight vector \( \mathbf{w}_{Entropy} \).

\begin{table}[h!]
\caption{Weighted decision matrix for RSSI use case (Entropy-TOPSIS)}
\label{tab:uc1-weighted}
\centering
\begin{tabular}{l|ccc}
\toprule
Candidate & RSSI & Load & Delay \\
\midrule
gNB & 0.04756 & 0.69803 & 0.00797 \\
AP5 & 0.01893 & 0.00000 & 0.00728 \\
AP7 & 0.05988 & 0.15512 & 0.00523 \\
\bottomrule
\end{tabular}
\end{table}

The computed weighted decision matrix is then used to rank the candidate nodes according to the TOPSIS algorithm described in Section~\ref{sec_bg_mcdm_topsis}. The ideal and negative-ideal solutions for the weighted criteria are extracted from the best and worst values in each column using eqn~\eqref{eq:ideal_solutions}. It is to be noted that RSSI is the a benefit criterion while load and delay are cost criteria. Table~\ref{tab:uc1-ideal} shows the computed ideal and negative-ideal solutions.

\begin{table}[h!]
\caption{Ideal and negative-ideal solutions for RSSI use case (Entropy-TOPSIS)}
\label{tab:uc1-ideal}
\centering
\begin{tabular}{l|cc}
\toprule
Criterion & \( v_j^+ \) (Ideal) & \( v_j^- \) (Negative-Ideal) \\
\midrule
RSSI  & 0.05988 & 0.01893 \\
Load  & 0.69803 & 0.00000 \\
Delay & 0.00797 & 0.00523 \\
\bottomrule
\end{tabular}
\end{table}

The next step involves computing the Euclidean distances from the ideal and negative-ideal solutions using eqn~\eqref{eq:distance_idealsols}. These are reported in Table~\ref{tab:uc1-dist}.

\begin{table}[h!]
\caption{Distance to ideal and negative-ideal solutions (RSSI use case (Entropy-TOPSIS))}
\label{tab:uc1-dist}
\centering
\begin{tabular}{l|cc}
\toprule
Candidate & \( S_i^+ \) & \( S_i^- \) \\
\midrule
gNB & 0.69814 & 0.02863 \\
AP5 & 0.04099 & 0.69803 \\
AP7 & 0.15512 & 0.54446 \\
\bottomrule
\end{tabular}
\end{table}

Finally, the relative closeness to the ideal solution is computed using eqn~\eqref{eq:closeness_coeff}. The closeness values and resulting rankings for each alternative are presented in Table~\ref{tab:uc1-closeness}.

\begin{table}[h!]
\caption{Relative closeness and final ranking for RSSI use case (Entropy-TOPSIS)}
\label{tab:uc1-closeness}
\centering
\begin{tabular}{l|cc}
\toprule
Candidate & \( C_i^* \) & Rank \\
\midrule
gNB & 0.03939 & 3 \\
AP5 & 0.94453 & 1 \\
AP7 & 0.77827 & 2 \\
\bottomrule
\end{tabular}
\end{table}

\paragraph{AHP based TOPSIS}\label{sec_perfAnalysis_usecase1_ahp} For the case of AHP based TOPSIS, we first pre-compute the AHP based weights, and then rank the candidate nodes according to the TOPSIS algorithm. Unlike the case of TOPSIS, where weights are computed dynamically based on the instantaneous decision matrix, AHP uses precomputed weights according to the pairwise comparisons. So the weighing process does not repeat at every decision making instance. The process begins with expert judgment to construct the pairwise comparison matrix, as defined in Section~\ref{sec_bg_mcdm_ahp}. Generally, as the connection integrity depends on RSSI, in this use case, it is considered as the priority and followed by access-node load and link delay respectively. The pairwise comparison matrix  $\textbf{P}$, that quantifies this order of priority is given in Table~\ref{pairwise_1}.  $\textbf{P}$ is then normalized, using eqn~\eqref{eq:ahp_pairwise_norm}, followed by computation of weights \( \mathbf{w}_{AHP} \) using eqn~\eqref{eq:ahp_weight}. These weights are shown in Table~\ref{tab:uc1-ahp-weights}.

\begin{table}[h]
\caption{Pairwise Comparison Matrix for AHP}
\label{pairwise_1}
\centering
\begin{tabular}{llll}
\hline
      & RSSI & Load & Delay \\ \hline
RSSI  & 1    & 2    & 4     \\
Load  & 0.5  & 1    & 3     \\
Delay & 0.25 & 0.33 & 1     \\ \hline
\end{tabular}
\end{table}

\begin{table}[h!]
\caption{AHP based weights for RSSI use case}
\label{tab:uc1-ahp-weights}
\centering
\begin{tabular}{l|c}
\toprule
Criterion & AHP based weight \( \mathbf{w}_{AHP} \) \\
\midrule
RSSI      & 0.55714 \\
Load      & 0.32024 \\
Delay     & 0.12262 \\
\bottomrule
\end{tabular}
\end{table}

Using the \( \mathbf{w}_{AHP} \), the TOPSIS method described in Section~\ref{sec_bg_mcdm_topsis} is followed to obtain the AHP-based MCDM method's rankings as follows. First, the decision matrix described for this use case, as shown in Table~\ref{tab:uc1-raw} is normalized according to vector normalization eqn~\eqref{eq:topsis_vector_normalization}, to obtain the weighted normalized decision matrix. The normalized decision matrix is then multiplied with the \( \mathbf{w}_{AHP} \) vector, using eqn~\eqref{eq:norm_weighted_matrix} to get the weighted normalized decision matrix, shown in Table~\ref{tab:uc1-ahp-weighted}.

\begin{table}[h!]
\caption{Weighted normalized decision matrix for RSSI use case (AHP-TOPSIS)}
\label{tab:uc1-ahp-weighted}
\centering
\begin{tabular}{l|ccc}
\toprule
Candidate & RSSI & Load & Delay \\
\midrule
gNB & 0.33637 & 0.31261 & 0.08143 \\
AP5 & 0.13391 & 0.00000 & 0.07445 \\
AP7 & 0.42347 & 0.06947 & 0.05351 \\
\bottomrule
\end{tabular}
\end{table}

Next, the ideal and negative-ideal solutions are computed from the weighted normalized decision matrix using eqn~\eqref{eq:ideal_solutions} and presented in Table~\ref{tab:uc1-ahp-ideal}. It is to be noted that RSSI is a benefit criterion while access-node load and link delay are cost criteria.

\begin{table}[h!]
\caption{Ideal and negative-ideal solutions for RSSI use case (AHP-TOPSIS)}
\label{tab:uc1-ahp-ideal}
\centering
\begin{tabular}{l|cc}
\toprule
Criterion & \( v_j^+ \) (Ideal) & \( v_j^- \) (Negative-Ideal) \\
\midrule
RSSI  & 0.42347 & 0.13391 \\
Load  & 0.00000 & 0.31261 \\
Delay & 0.05351 & 0.08143 \\
\bottomrule
\end{tabular}
\end{table}

Next, the distance of each candidate from the ideal and negative-ideal solutions is calculated using eqn~\eqref{eq:distance_idealsols} and shown in Table~\ref{tab:uc1-ahp-dist}.

\begin{table}[h!]
\caption{Distances to ideal and negative-ideal solutions for RSSI use case (AHP-TOPSIS)}
\label{tab:uc1-ahp-dist}
\centering
\begin{tabular}{l|cc}
\toprule
Candidate & \( S_i^+ \) & \( S_i^- \) \\
\midrule
gNB & 0.32572 & 0.20246 \\
AP5 & 0.29031 & 0.31269 \\
AP7 & 0.06947 & 0.37913 \\
\bottomrule
\end{tabular}
\end{table}

Finally, the relative closeness \( C_i^* \) is computed using ~eqn\eqref{eq:closeness_coeff}, and the final ranking is summarized in Table~\ref{tab:uc1-ahp-closeness}.

\begin{table}[h!]
\caption{Relative closeness and ranking for RSSI use case (AHP-TOPSIS)}
\label{tab:uc1-ahp-closeness}
\centering
\begin{tabular}{l|cc}
\toprule
Candidate & \( C_i^* \) & Rank \\
\midrule
gNB & 0.38332 & 3 \\
AP5 & 0.51855 & 2 \\
AP7 & 0.84514 & 1 \\
\bottomrule
\end{tabular}
\end{table}

\paragraph{Comparison of AHP-TOPSIS and Entropy-TOPSIS for RSSI prioritized use case:}
\begin{figure}[htbp]
\centerline{\includegraphics[width=\columnwidth]{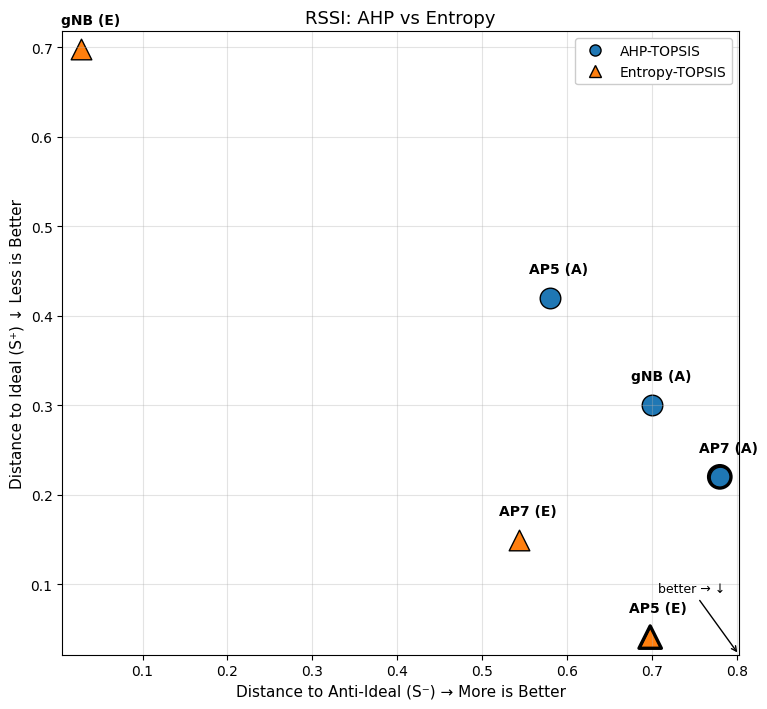}}
\caption{TOPSIS based distance visualization for RSSI use case.}
\label{fig:uc1-topsis-vis}
\end{figure}
In RSSI use case, AHP-TOPSIS ranks AP7 first, while Entropy-TOPSIS prefers AP5. This distinction is rooted in the weighting approach. AHP uses fixed weights prioritizing RSSI, leading to AP7—which has the strongest signal—being ranked highest. Conversely, entropy-based weighting emphasizes the criterion with the greatest variability, which in this case is access-node load. As a result, AP5, the access point with the lowest load, is selected.

Figure~\ref{fig:uc1-topsis-vis} provides a visual comparison of the AHP-TOPSIS and Entropy-TOPSIS results for RSSI use case. Each point represents an access point, plotted by its distance to the ideal solution ($S^+$, to be minimized) and to the anti-ideal solution ($S^-$, to be maximized). Points closer to the bottom-right are considered better. Circles indicate results from AHP-TOPSIS, and squares represent Entropy-TOPSIS. The labels include “(A)” for AHP and “(E)” for Entropy. The color of each point reflects its distance from the anti-ideal solution — greener points are better, meaning farther from the worst-case. The plot clearly shows that AP7 is ranked highest under AHP-TOPSIS, while AP5 is preferred by Entropy-TOPSIS, illustrating how weighting methods influence the final decision.

This use case demonstrates that Entropy-TOPSIS excels when real-time data variation is meaningful and load-aware decisions are critical. On the other hand, AHP-TOPSIS offers a more stable, preference-driven approach—useful when network variability may obscure true performance differences. 

\subsection{ UEs Near-gNB — Load-Prioritized Use Case}\label{sec_perfAnalysis_usecase2} When UEs are close to the 5G gNB, the received signal strength from the gNB is typically much higher than from nearby WiFi APs because of the gNB’s higher transmit power. At the same time, the gNB tends to be more heavily loaded than the APs. In such scenarios, the MCDM algorithms, which treat RSSI as a benefit criterion, are often drawn toward the gNB; when RSSI dominates, both Entropy-TOPSIS and AHP-TOPSIS can rank the gNB first despite congestion. This is illustrated with a UE within range of two WiFi APs and one gNB under the parameter constraints stated above.

Table~\ref{tab:uc2-raw} gives the decision matrix formed by considering example RSSI, access-node load, and link delay values for the three alternatives.

\begin{table}[h!]
\caption{Raw decision matrix for load use case}
\label{tab:uc2-raw}
\centering
\begin{tabular}{l|ccc}
\toprule
Candidate & RSSI (dB) & Load & Delay (sec) \\
\midrule
gNB & -62 & 29 & 0.045 \\
AP5 & -72 & 3  & 0.025 \\
AP7 & -85 & 1  & 0.023 \\
\bottomrule
\end{tabular}

\end{table}

In a similar manner as Section~\ref{sec_perfAnalysis_usecase1}, by following the steps mentioned in Section~\ref{sec_bg_mcdm} and Eqs.~(\ref{eq:ahp_pairwise_norm})--(\ref{eq:closeness_coeff}), we calculate the ranking outcomes for this use case using Entropy–TOPSIS and AHP–TOPSIS. Table~\ref{tab:combined-rankings-ENTROPY-AHP-UC2} gives the relative closeness scores and rankings of the nodes for the MCDM method. It can be seen that both both Entropy-TOPSIS and AHP-TOPSIS select the gNB, even though its load is much higher. The gNB’s dominant RSSI outweighs the load and delay disadvantages, which in turn exacerbates congestion on the cellular side, despite AP5 offering sufficient RSSI with light load and thus being a viable offloading target.

\begin{table}[h!]
\caption{Comparison of relative closeness scores (\( C_i^* \))  and rankings across Entropy-TOPSIS AND AHP-TOPSIS for Load Use Case}
\label{tab:combined-rankings-ENTROPY-AHP-UC2}
\centering
\begin{tabular}{l|cc|cc}
\toprule
Candidate & \multicolumn{2}{c|}{Entropy-TOPSIS} & \multicolumn{2}{c|}{AHP-TOPSIS} \\
                      & \( C_i^* \) & Rank & \( C_i^* \) & Rank \\
\midrule
gNB & 0.5173 & 1 & 0.6346 & 1 \\
AP5 & 0.4845 & 2 & 0.3688 & 2 \\
AP7 & 0.4827 & 3 & 0.3654 & 3 \\
\bottomrule
\end{tabular}

\end{table}

To make AHP-TOPSIS suitable for such congestion scenarios, the pairwise matrix used to obtain the AHP weights can be modified such that the load gets higher priority compared to RSSI and Delay. However, this comes at the risk of selecting a node with insufficient RSSI due to the higher priority on load. We show that by modifying the preference of the criteria, AHP can be applicable in such scenarios where Entropy based TOPSIS fails. This modified AHP-TOPSIS is labeled as $ AHP(L)$. The modified pairwise comparison matrix is shown in Table~\ref{pairwise_2}. The weights corresponding to the $ AHP(L)$, \( \mathbf{w}_{AHP(L)} \) are obtained using eqn~\eqref{eq:ahp_weight} and are provided in Table~\ref{tab:AHP(L)_weights}

\begin{table}[h]
\caption{Pairwise Comparison Matrix for AHP(L)}
\label{pairwise_2}
\centering
\begin{tabular}{llll}
\hline
      & RSSI & Load & Delay \\ \hline
RSSI  & 1    & 0.5   & 4     \\
Load  & 2  & 1    & 3     \\
Delay & 0.25 & 0.33 & 1     \\ \hline
\end{tabular}

\end{table}

\begin{table}[h!]
\caption{Weights using AHP(L)}
\label{tab:AHP(L)_weights}
\centering
\begin{tabular}{l|c}
\toprule
Criterion & AHP(L) based weight \( \mathbf{w}_{AHP(L)} \) \\
\midrule
RSSI    &  0.36 \\
Load    &  0.51 \\
Delay   &  0.13 \\
\bottomrule
\end{tabular}
\end{table}

Using the weights, \( \mathbf{w}_{AHP(L)} \), the TOPSIS algorithm is repeated using the procedure mentioned in Section~\ref{sec_bg_mcdm_topsis} and Eqs.~(\eqref{eq:topsis_vector_normalization} -- \eqref{eq:closeness_coeff}), to obtain the relative closeness scores and rankings. Table~\ref{tab:combined-rankings-ENTROPY-AHPL-UC2} compares the relative closeness scores (\( C_i^* \))  and rankings across Entropy-TOPSIS and AHP(L)-TOPSIS. It can be seen that AHP(L) elevates AP5 to the top rank and places the gNB last, enabling offloading to a feasible AP and reducing cellular load.

\begin{table}[!t]
\caption{Comparison of relative closeness scores (\( C_i^* \))  and rankings across Entropy-TOPSIS and AHP(L)-TOPSIS}
\label{tab:combined-rankings-ENTROPY-AHPL-UC2}
\begingroup
\setlength{\belowcaptionskip}{6pt} 
\centering
\begin{tabular}{l|cc|cc}
\toprule
Candidate & \multicolumn{2}{c|}{Entropy-TOPSIS} & \multicolumn{2}{c}{AHP(L)-TOPSIS} \\
                      & \( C_i^* \) & Rank & \( C_i^* \) & Rank \\
\midrule
gNB & 0.5173 & 1 & 0.4142 & 3 \\
AP5 & 0.4845 & 2 & 0.5861 & 1 \\
AP7 & 0.4827 & 3 & 0.5858 & 2 \\
\bottomrule
\end{tabular}
\endgroup
\end{table}

\begin{figure*}[htbp]
\centerline{\includegraphics[width=2\columnwidth]{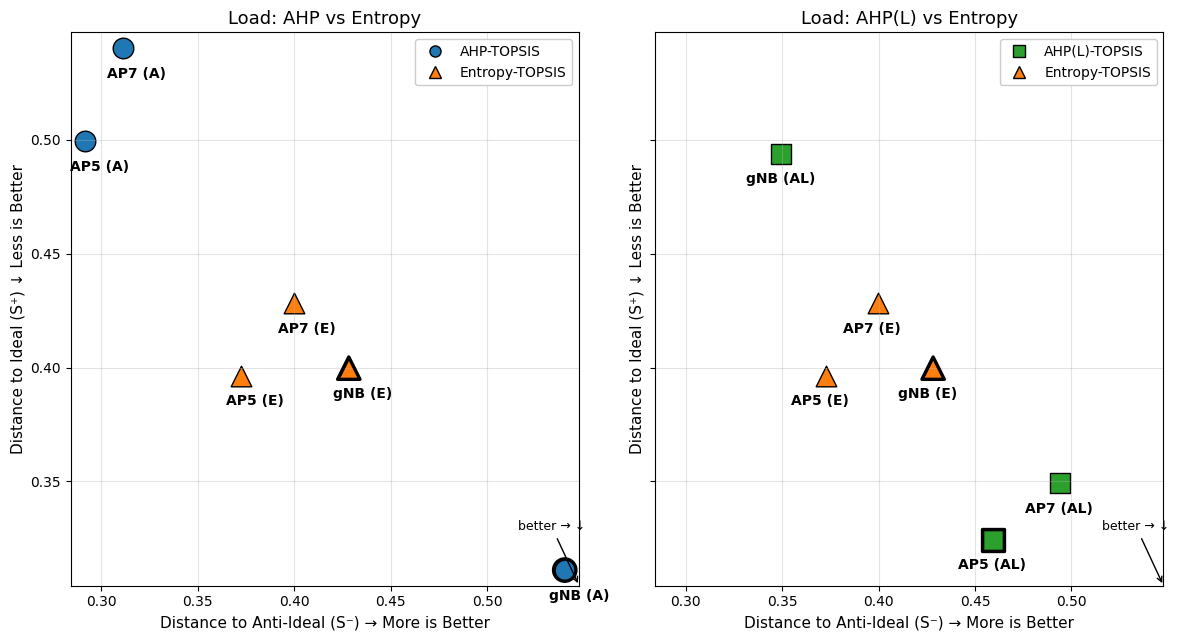}}
\caption{TOPSIS based distance visualization for Load use case.}
\label{fig:uc2-topsis-vis}
\end{figure*}

\paragraph{Comparison of AHP-TOPSIS and Entropy-TOPSIS for Load use case:}

In the Load Priority use case, the UE is positioned close to the gNB, resulting in disproportionately higher RSSI at the gNB than at nearby WiFi APs. Figure~\ref{fig:uc2-topsis-vis} visualizes the results for Entropy-TOPSIS vs AHP-TOPSIS, and Entropy-TOPSIS vs AHP(L)-TOPSIS (Load priority). Both Entropy-TOPSIS and AHP-TOPSIS rank gNB first, followed by AP5 and AP7. Entropy-TOPSIS, being purely data-driven, selects gNB due to its dominant RSSI despite its high load. Similarly, AHP-TOPSIS with RSSI priority favors gNB as it explicitly prioritizes signal strength in the decision. By contrast, AHP(L)-TOPSIS (operator-biased AHP with load prioritized) ranks AP5 first and the gNB last. This switch demonstrates that if weights are manually re-tuned to prioritize access-node load over RSSI, AHP can counteract selection bias toward the gNB and offload effectively under congestion. However, this comes with two practical drawbacks. First, such re-weighting requires prior knowledge or manual detection of congestion to select the appropriate weight set, and thus does not generalize across scenes without continual retuning. Second, prioritizing access-node load this way can sacrifice link reliability, because the AP that is least loaded may have insufficient RSSI to sustain a robust post-handover connection (i.e., below the operational RSSI threshold).

\subsection{Conclusion} Overall, while Entropy-TOPSIS may be unbiased and AHP can be adjusted to emphasize access-node load, these MCDM methods—used directly—do not consistently realize multi-RAT offloading under uncertain network conditions. A method-agnostic offloading guard is therefore required to (i) prevent gNB over-selection when its load is high and (ii) enforce feasibility (e.g., minimum RSSI/range) so that offloading does not select an AP that cannot sustain the link.

\section{Congestion-aware Offloading and Handover via Empirical RAT Evaluation - \chrome{}} \label{sec_ho_algo}

In this section, we describe the proposed \chrome{} framework for dense multi-RAT networks. The objective of this work is to provide a generalized MCDM-based handover and offloading framework for heterogeneous multi-RAT networks that supports robust, policy-aligned decisions across technologies with diverse performance characteristics. Conventional RSSI based handover transfers the link to the node with the highest received signal strength. In heterogeneous deployments, such as the model described in Section~\ref{sec_architecture}, higher gNB transmit power from the macro cell can dominate RSSI and obscure viable WiFi APs that offer comparable QoS. This bias toward high-power links limits the ability to account for access-node load and latency conditions during selection. We therefore move beyond signal strength and jointly consider RSSI, link delay, and access-node load. These criteria capturing physical proximity, signal strength, latency (delay), and node utilization/congestion state (load), enabling congestion-aware offloading to suitable alternatives when high-power links are overloaded.

The evaluation of the two use cases in Section~\ref{sec_mcdmusecases} show that joint utilization of these criteria in the handover decision phase is possible through the use of MCDM methods. We also summarize how the subjective and objective MCDM weighting schemes—AHP-TOPSIS and Entropy-TOPSIS—exhibit distinct decision behaviors. AHP-TOPSIS encodes handover goals via pairwise criterion preferences, whereas Entropy-TOPSIS adapts to the most discriminative criteria at each decision epoch, reflecting real-time network conditions. As discussed in Section~\ref{sec_bg_mcdm_weighting_compare}, these properties make the two approaches suitable for different scenarios. However, in congested settings such as Section~\ref{sec_perfAnalysis_usecase2}, both schemes can fail to prefer the node with lower load and instead reselect the congested one. Entropy-TOPSIS may overweight RSSI when signal dispersion dominates, while AHP-TOPSIS requires manual re-tuning (e.g., load-first) that presumes prior knowledge and can compromise stability when RSSI is marginal. 

To address this, we propose \chrome{}: a framework that retains MCDM as the scoring backbone but inserts a method-agnostic, RAT-based RSSI threshold before the final node selection. The threshold guards against the top-ranked node being the overloaded node and redirects traffic to the next-best alternative. Combined with the dual-weighting pipeline, this yields stable, policy-aligned offloading in multi-RAT networks.

Building on our earlier work (entropy-based TOPSIS with two criteria: RSSI and delay), we extend the design in two directions. First, we include the load on a node alongside RSSI and link delay so that the decision captures a node's utilization/congestion state. It enables to directly optimize the high congestion scenarios. Second, we integrate both AHP- and entropy-based weighting within the same pipeline. This allows operation under the different scenarios identified in Sections~\ref{sec_bg_mcdm_weighting_compare} and~\ref{sec_perfAnalysis_usecase2}, while the guard prevents re-selection of congested nodes even when the score favors them.

\subsection{Algorithm Description: \chrome{} for multi-RAT handover and offloading}

\begin{figure}[htbp]
\centerline{\includegraphics[width=\columnwidth]{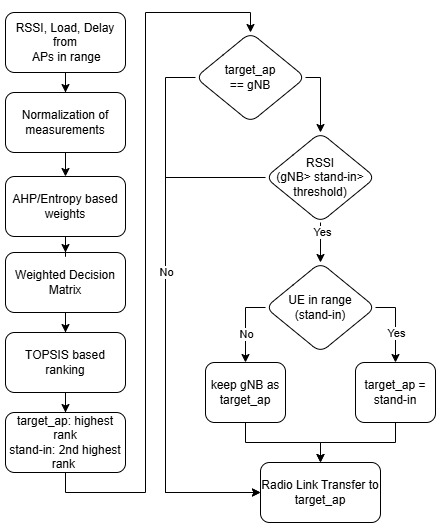}}
\caption{The proposed \chrome{} framework with two types of weighing and RAT-based RSSI threshold}
\label{HO_mcdm_algo}
\end{figure}
The proposed \chrome{} framework is illustrated in Figure~\ref{HO_mcdm_algo}. We first enumerate all candidate APs within range of the UE. If the UE lies within the gNB’s coverage, the gNB is included as a candidate. The corresponding RSSI of each UE with respect to a node(AP/gNB), instantaneous load on the associated node and instantaneous link delay of the UE-node link is collected, to form the decision matrix. These measurements form the decision matrix, which is then normalized.

Following Section~\ref{sec_bg_mcdm}, weights are obtained and alternatives are ranked. For AHP-based weighting, we use the predetermined values in Table~\ref{tab:uc1-ahp-weights} (as instantiated in Section~\ref{sec_perfAnalysis_usecase1}); for entropy-based weighting, we compute entropy from the normalized matrix, derive the criterion divergences, and obtain weights as described in Section~\ref{sec_bg_mcdm_entropy}. (These steps mirror those in Section~\ref{sec_perfAnalysis_usecase1} for AHP-TOPSIS and Entropy-TOPSIS.) Multiplying the weight vector with the normalized matrix yields the weighted decision matrix. We then apply the TOPSIS procedure (Section~\ref{sec_bg_mcdm_topsis}) to rank the candidates by proximity to the ideal and distance from the negative-ideal solutions. This completes the MCDM ranking stage.

\emph{RAT-based RSSI threshold:} Using the TOPSIS rankings, the highest-ranked candidate is designated the \emph{target} and the second-ranked candidate the \emph{stand-in}. To enable congestion-aware offloading without sacrificing feasibility, we apply a RAT-based RSSI threshold at selection time: if the target is the gNB, we examine the the RSSI value of the stand-in. when the stand-in is a WiFi AP and its RSSI exceeds a predefined threshold (e.g., $-85$\,dBm) and the UE is within coverage, we redirect the handover to the stand-in AP. Finally, the radio link is transferred to the selected node. 

This selection guard promotes offloading to a qualifying alternative under congestion, while maintaining QoS by combining RSSI (signal strength), link delay (user-experienced latency), and access-node load (congestion state) in the score, and enforcing a per-RAT signal floor at selection, the procedure supports stable, congestion-aware offloading decisions in heterogeneous multi-RAT deployments.

\section{Multi-RAT network simulation} \label{sec_perfAnalysis_sim}

The architecture described in Section~\ref{sec_architecture} is realized using the Mininet WiFi emulator \cite{mininetwifi} with a RYU controller \cite{ryu}. The SDN RYU controller is an open-source Python-based SDN controller used to run an L3 switch, which implements an IP-based packet matching. Mininet WiFi is an extension of the Mininet emulator \cite{mininet}, which is a widely used tool for wired SDN research. Mininet WiFi provides support for wireless integration of the SDN controller with the WiFi access points through OpenFlow-enabled switches. There is no inherent support for cellular technologies such as LTE and 5G NR, so custom modifications were made to Mininet WiFi to create nodes similar to gNBs. Table~\ref{network_setup} gives the network parameters used for each RAT. The MCS index of all the nodes is set to 12, with the maximum transmit bit rate as 78.0 MBps. For gNBs the frequency of operation was set to 2.412GHz, and the transmit power was calculated using 3GPP's Non-Line of Sight Urban Macro Model for 5G NR \cite{5gnrchannelmodel}. Mininet WiFi uses generic propagation models that use path-loss exponents to calculate the RSSI at the receiver. So, the calculated transmit power and range using the 5G NR standards are mapped to the range of possible values within the urban mobility model \cite{goldsmith2005wireless}, and the Log Normal propagation model with a path loss exponent of 3.5 is selected. In the case of the WiFi APs, the frequency of operation is set to 5GHz, and the 802.11 ax standard is used.

\begin{table}[h]
\centering
    \caption{Mininet WiFi parameters for Multi RAT network simulation} \label{network_setup} 
    \resizebox{0.35\textwidth}{!}{%
    \begin{tabular}{c c c} 
    \toprule
    Parameter & WiFi AP & 5G BS\\    
    \midrule
    Frequency & 5.18-5.825GHz & 2.412GHz\\
    \midrule
    TxPower & 2dBm & 16dBm\\
    \midrule
    Range & 60m & 150m\\
    \midrule
    Mode & ax5 & g\\
    \midrule
    Propagation Model & LogNormal & LogNormal \\
    \midrule
    Pathloss exponent & 3.5 & 3.5 \\

\bottomrule
 \end{tabular}}
\end{table}

\begin{figure}[ht]
\centerline{\includegraphics[width=\columnwidth]{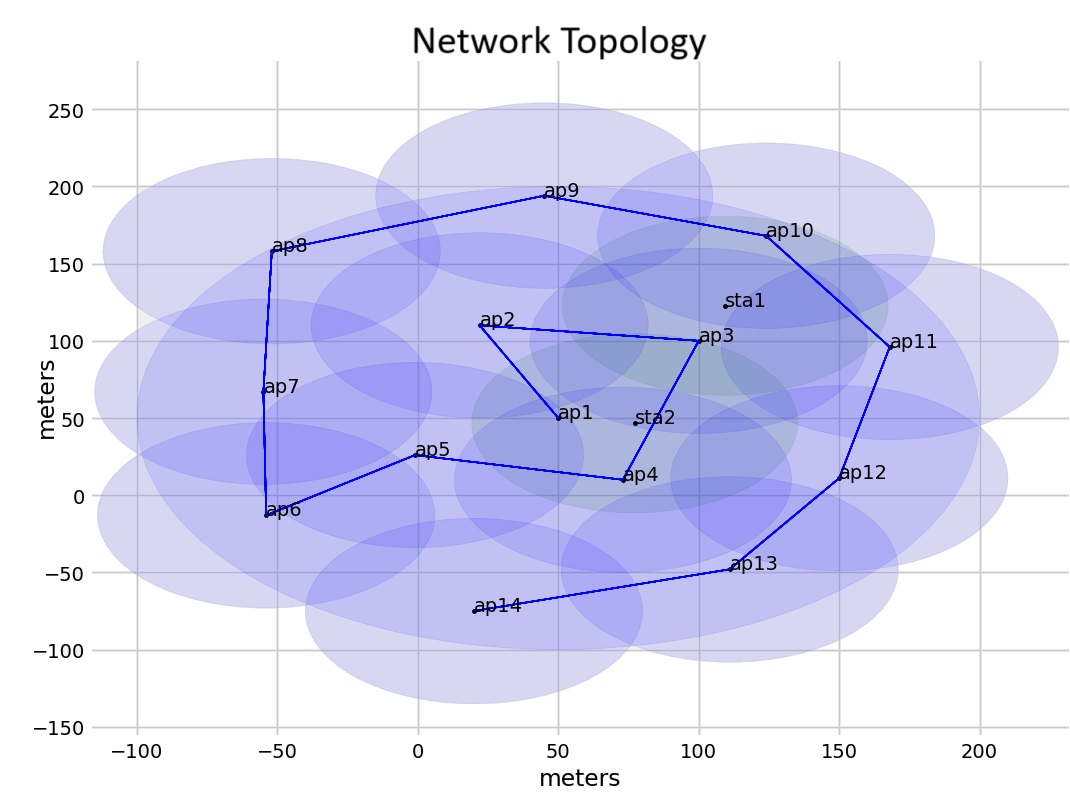}}
\caption{Simulated Multi RAT network with 2 mobile UEs}
\label{network_topo}
\end{figure}

The topology consists of one gNB and 14 APs, with the APs evenly distributed across the coverage area of the gNB. An increasing number of UEs is distributed randomly across the topology. Each UE follows the "random direction" mobility model, defined in Mininet WiFi. Each simulation run is carried out with 16, 32, 64 UEs to mimic a densely populated, scalable urban environment. Fig. \ref{network_topo} shows the simulated network topology with 2 UEs denoted as 'sta', gNB at the center of the large coverage area, denoted by 'gNB', and 14 WiFi APs, denoted as 'AP2-AP15'. A hard handover is used in Mininet WiFi, i.e., the connection to the existing node is dropped before the radio link is transferred to the target node. Using this topology, the simulation is carried out with the proposed \chrome{} framework for (i) AHP based TOPSIS, (ii) entropy based TOPSIS, as described in Section~\ref{sec_ho_algo}, and (iii) a conventional RSSI-based handover,  for a predetermined time duration, 60 minutes. Each algorithm is run 15 times and the averages across these runs are collected for each performance metric.

The objective of the proposed \chrome{} framework in Section~\ref{sec_ho_algo} is to facilitate the handover and offloading of users in a multi-RAT network while improving the QoS by reducing the delay experienced by the users, reducing the load on the 5G network and maintaining the throughput. To achieve this, the proposed framework considers RSSI, access-node load, and link delay as explained in Section~\ref{sec_ho_algo}. A multi-RAT system is simulated using Mininet WiFi, and the following parameters are collected to verify the described objectives. 
\begin{itemize}
    \item Load on the 5G network: The number of stations connected to the 5G gNB after every handover attempt. These values are collected from all nodes in the network, but we show the load on the 5G gNB to evaluate the impact of the proposed handover and offloading mechanism. 
    \item Link Delay Cost: The link delay cost, used to evaluate handover performance, is defined as the difference in average delay before and after the handover event. The average link delay, a key parameter in the MCDM framework, represents the typical round-trip time (RTT) for ICMP Echo Request packets traveling from a user device to a destination and back. It reflects network latency and is computed by averaging the RTTs of all transmitted packets. So, the link delay cost is obtained by subtracting the previous node’s average delay from that of the new target node. A negative delay cost implies an improvement in latency following the handover, indicating a more responsive connection. Conversely, a positive value suggests that the new connection point introduces additional delay, representing a degradation in service quality.
    \item Throughput: The amount of data transferred over a link in a given time is collected after each handover attempt. (A failed handover results in approximately zero throughput.) The values are collected across the network simulation time, and the cumulative throughput is used to compare the MCDM and conventional handovers.
    \item Number of handover attempts: The number of handovers reflects the effectiveness of the proposed \chrome{} framework in reducing total handovers, while reducing the access-node load. Additionally, the we also provide the number of WiFi to 5G handovers, which are directly responsible for increasing the load on 5G network. An effective offloading approach would reduce the load on the 5G network, while having lower number of total handovers and lower number of WiFi to 5G handover.

\end{itemize}

\subsection{Numerical Results}
\label{sec_perfAnalysis_results}

\begin{figure}[h]
\centerline{\includegraphics[width=\columnwidth]{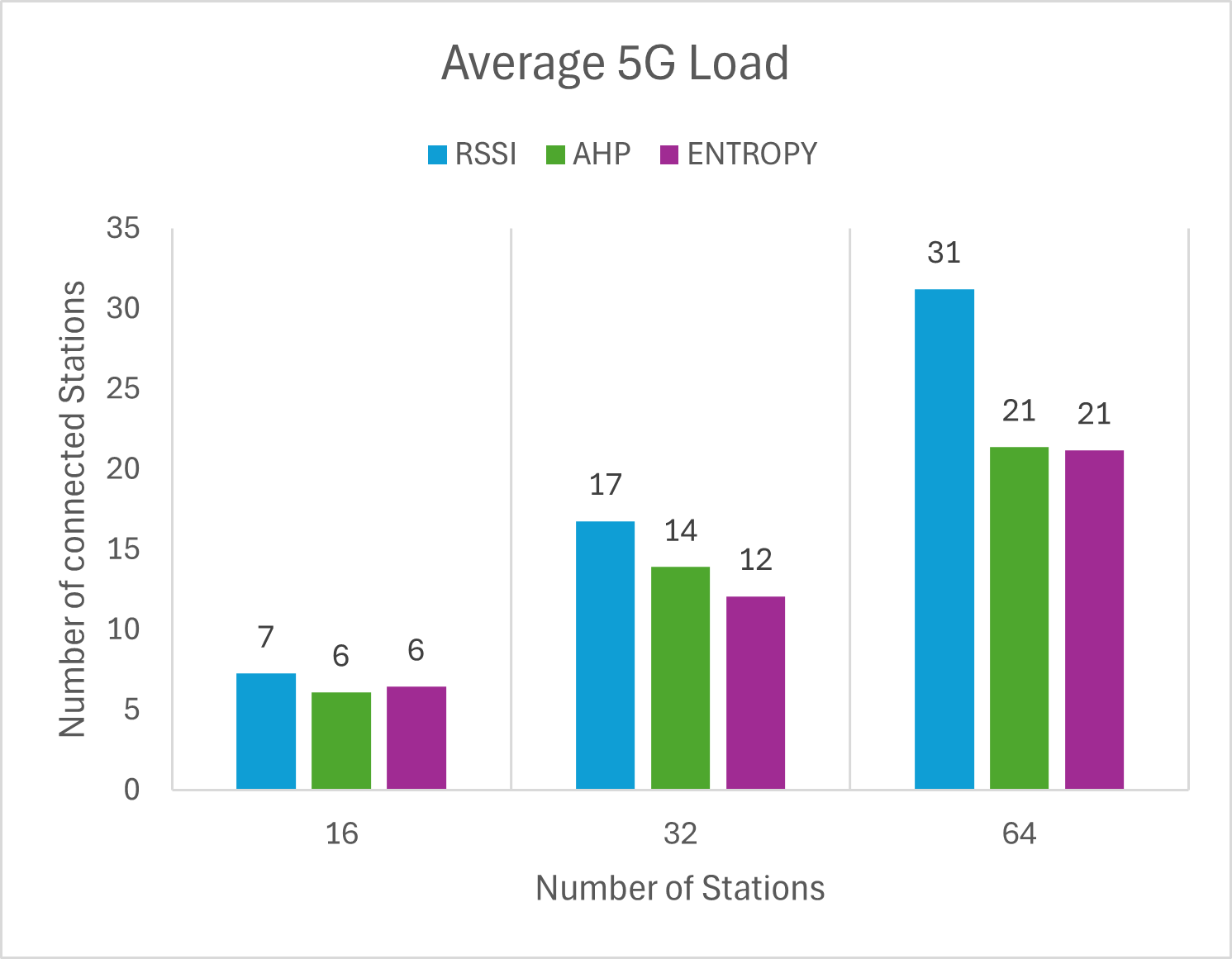}}
\caption{Average 5G load for 16, 32, 64 stations}
\label{load}
\end{figure}

\begin{figure}[!t]
\centering
\subfloat[Total Number of Handovers\label{num_totalHO}]{
  \includegraphics[width=\columnwidth]{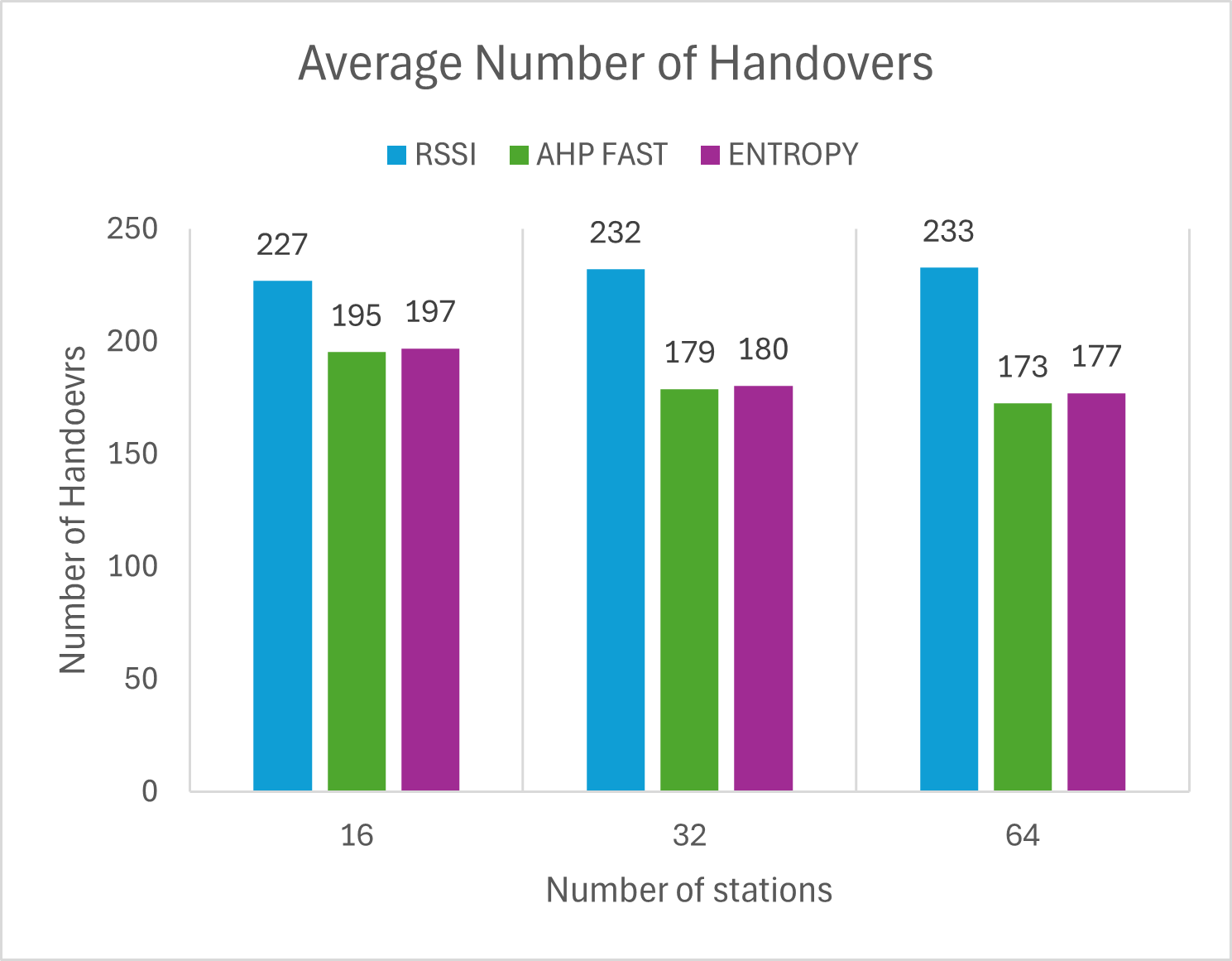}
}
\hfill
\subfloat[Number of WiFi to 5G Handovers\label{num_WiFi25gHO}]{
  \includegraphics[width=\columnwidth]{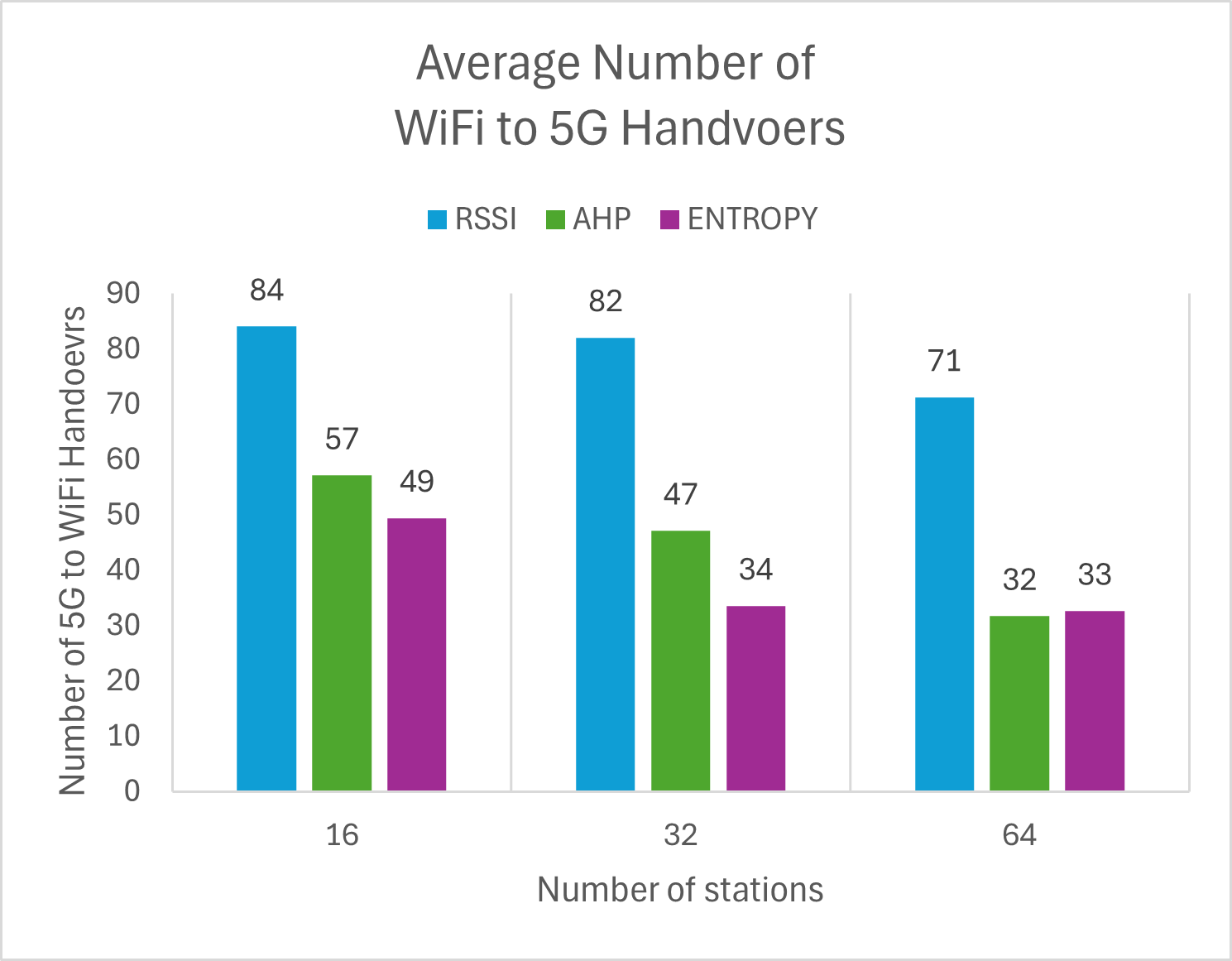}
}
\caption{Comparison of total handovers and WiFi$\to$5G handovers for 16, 32, and 64 stations.}
\label{Num_HOs}
\end{figure}

Fig. \ref{load},\ref{Num_HOs},\ref{delay},\ref{throughput} show the performance of the proposed \chrome{} framework using AHP and Entropy based weights, for an increasing number of UEs in the network. RSSI refers to the conventional RSSI-based handover, AHP refers to the \chrome{} framework using AHP based weights, Entropy refers to the proposed \chrome{} framework with Entropy-based weights. The results show that the proposed \chrome{} framework with both the weighing methods, performs better than the conventional method for all three cases, across all performance metrics. A detailed analysis of each performance metric is as follows:

\subsubsection{Load on the 5G network} As shown in the figure Figure~\ref{load}, load on the 5G network is highest in case of the RSSI based handover. Both MCDM-based methods AHP and Entropy significantly reduce the load on the 5G network compared to the traditional RSSI-based handover approach. For a scenario with 16 stations, AHP based MCDM offloading achieves a 16\% reduction in 5G gNB load, while Entropy based MCDM offloading results in an 11\% reduction. When the number of stations increases to 32, AHP yields a 17\% reduction, and Entropy performs even better with a 28\% reduction. At 64 stations, the load reduction becomes more pronounced, with AHP and Entropy reducing 5G usage by 31\% and 32\%, respectively. With increasing number of stations, both the MCDM based offloading methods perform better in reducing the load on the 5G gNB. This trend highlights a critical limitation of the RSSI-based approach: as the number of stations increases, more devices tend to connect to the 5G network due to its inherently higher RSSI levels, attributed to the higher transmit power of 5G base stations, as discussed in Section~\ref{sec_ho_algo}. However, this often leads to congestion and degraded QoS. In contrast, the proposed AHP and entropy based \chrome{} framework strategies intelligently offload traffic to WiFi access points when they meet a minimum satisfactory RSSI threshold. By evaluating multiple criteria—such as signal strength, delay, and throughput—MCDM methods are able to identify the "next best" node rather than relying solely on signal strength. This enables effective multi-RAT handovers, distributing 5G gNB load more evenly and maintaining, or even enhancing, the overall QoS. As demonstrated in subsequent performance metrics, this approach not only alleviates 5G congestion but also contributes to reduced delay and comparable or improved throughput.

\subsubsection{Number of Handovers} A key performance performance indicator in evaluating handover strategies is the total number of handovers triggered by a given scheme. While successful handovers ensure continuity in connectivity, an excessive number of handovers, particularly rapid and repeated, can lead to the undesirable ping-pong effect degrading the user experience and network stability. Moreover, in offloading scenarios, frequent handovers to the 5G network not only burden it unnecessarily but also counteract the objective of traffic distribution. Thus, an optimal offloading centric handover algorithm should aim to minimize both the total handovers and those specifically directed toward the 5G infrastructure. Figure~\ref{Num_HOs} illustrates how the proposed \chrome{} framework compare with the traditional RSSI-based baseline. For the 16-station configuration, both AHP and Entropy reduce the overall number of handovers by approximately 13\%. With 32 stations, the reductions become more substantial, around 22\% for both methods. The trend continues with 64 stations, where both MCDM method have 24\% and 25\% fewer number of handovers respectively. These findings indicate that as the network becomes more populated, the AHP and Entropy methods are increasingly effective at reducing unnecessary handovers, offering more stable connectivity decisions. When focusing specifically on handovers to 5G, the MCDM methods consistently outperform the baseline by producing significantly fewer transitions. In the 16-station case, AHP results in 32\% fewer such handovers, while Entropy achieves a 41\% reduction. For 32 stations, these figures rise dramatically—43\% fewer for AHP and 60\% for Entropy. At 64 stations, both methods remain highly effective, with AHP and Entropy reducing WiFi-to-5G handovers by 55\% and 54\%, respectively. These results strongly support the efficiency of the proposed \chrome{} framework in not only balancing network load but also in reducing disruptive and unnecessary handovers to the 5G network.

\begin{figure}[h]
\centerline{\includegraphics[width=\columnwidth]{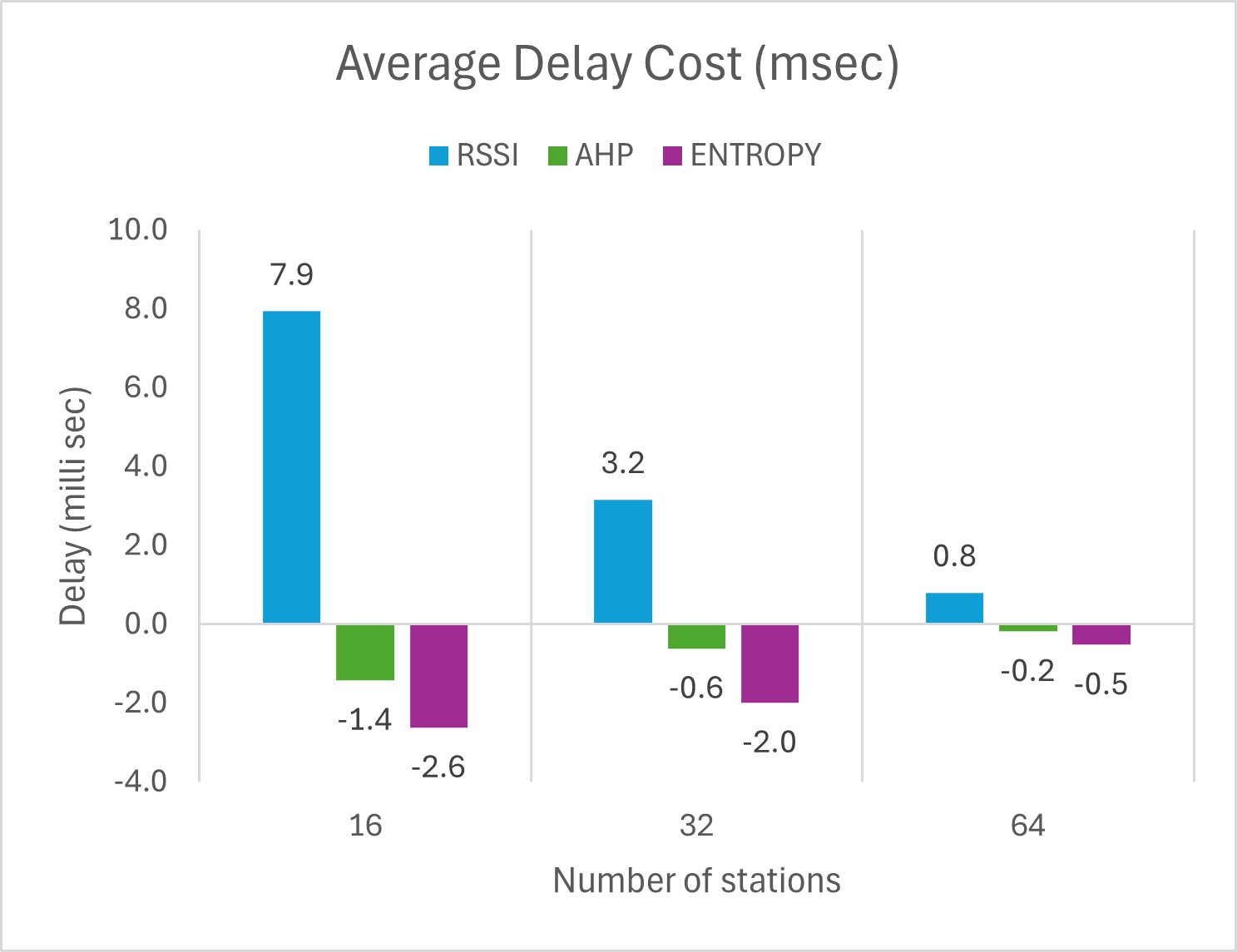}}
\caption{Delay Cost associated with the handover  for 16, 32, 64 stations}
\label{delay}
\end{figure}

\subsubsection{Delay Cost} The delay cost metric reflects the change in link delay experienced by users as a result of a handover. In traditional RSSI-based handovers, this metric tends to be positive, indicating that the delay increases after the handover, a sign of deteriorating link delay. This outcome is expected, as RSSI-only schemes do not account for link delay when making handover decisions. In contrast, MCDM-based methods, which incorporate delay as a decision criterion, are better equipped to maintain or improve QoS across heterogeneous network types, where latency characteristics can vary significantly. As demonstrated in the Fig.\ref{delay}, the RSSI-based baseline consistently shows a higher delay cost, confirming its lack of sensitivity to link quality during handovers. On the other hand, both AHP and Entropy-based MCDM methods exhibit negative delay costs, signifying an improvement in link performance following a handover. Quantitatively, the improvements are remarkable. For the 16-station scenario, AHP reduces delay cost by 118\%, while Entropy achieves a 133\% reduction compared to the RSSI baseline. In the 32-station case, delay cost improvements are 119\% and 163\% for AHP and Entropy, respectively. For the 64-station configuration, the reductions are even more pronounced—123\% for AHP and 166\% for Entropy respectively. These results clearly indicate that integrating integrating link delay into the handover process leads to significantly better latency outcomes, thereby enhancing the overall Quality of Service (QoS) delivered to the end user.

It can also be observed from Fig.~\ref{delay}, as the number of stations increases, the absolute delay cost for RSSI-based handovers decreases, while the delay cost for both AP and entropy based \chrome{} framework increases slightly. This trend arises not because RSSI decisions improve, but because the fixed set of access nodes becomes more actively utilized. With more stations connected, the average round-trip time (RTT) tends to decrease, as the link is used more efficiently and experiences less idle time. This behavior results in lower overall delay values for all methods as station density grows. Despite this reduction in absolute delay for RSSI, the percentage improvement delivered by MCDM methods remains substantial across all station counts. Even as delays drop overall, MCDM continues to make superior handover decisions by factoring in latency, achieving better outcomes than RSSI alone. This confirms the robustness of MCDM strategies in adapting to multi-RAT network conditions and optimizing handovers across both sparse and dense deployments.

\begin{figure}[h]
\centerline{\includegraphics[width=\columnwidth]{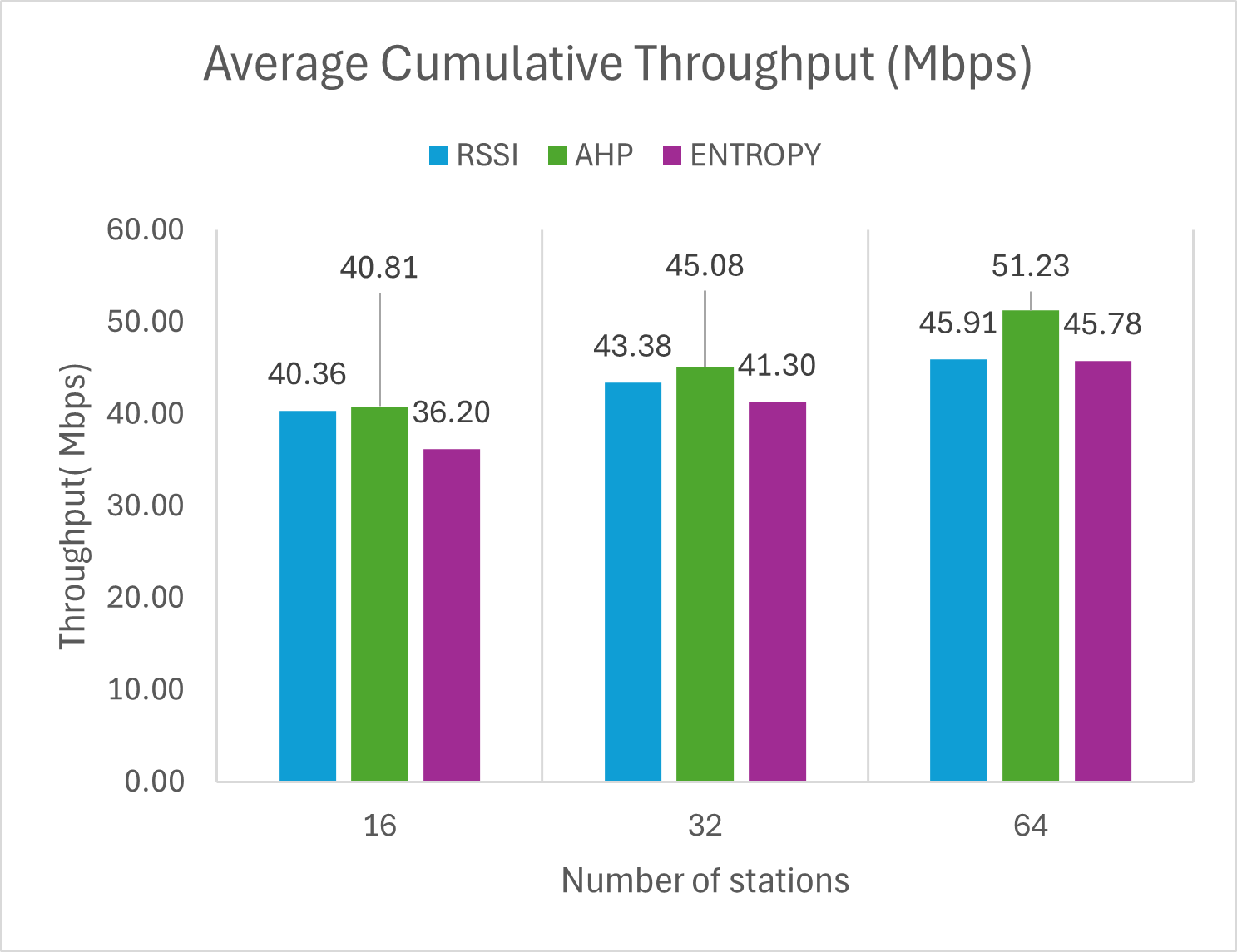}}
\caption{Average Cumulative Throughput for 16, 32, 64 stations}
\label{throughput}
\end{figure}

\subsubsection{Throughput}  

Fig. \ref{throughput} presents the throughput achieved under the three handover methods RSSI, AHP, and Entropy, as the number of stations increases. For 16 stations, AHP delivers 1\% more throughput than RSSI, while Entropy trails behind RSSI by 10\%. With 32 stations, AHP outperforms RSSI by 4\%, and Entropy provides 5\% lower throughput than RSSI-based handover, improving from the case of 16 stations. At 64 stations, AHP exceeds RSSI by 11\%, while Entropy achieves nearly identical throughput to the RSSI method.

These results underscore two important dynamics. First, AHP consistently outperforms RSSI across all station densities. As the number of stations increases, AHP’s performance advantage over RSSI becomes more pronounced. In contrast to RSSI-based methods, which often push users toward the strongest signal regardless of congestion, AHP balances signal quality with current network conditions. This balanced selection becomes increasingly advantageous as station density grows and network congestion intensifies. Avoiding overloaded nodes allows AHP to maintain higher throughput, especially in dense network scenarios. When many users connect to a single access point, bandwidth must be shared, resulting in lower per-user throughput. High loads also increase queuing delays, contention, and the likelihood of retransmissions, particularly in WiFi environments. By directing users to less congested nodes, AHP allows each device to access a larger share of available capacity, minimizing delays and maintaining efficient transmission, ultimately leading to higher throughput overall.

Entropy shows a different trend. In the case of lower stations, the variability in metrics like RSSI, delay, and access-node load is relatively random and inconsistent. Since the Entropy based weighing dynamically assigns a higher weight to metrics with greater variability, it may end up prioritizing access-node load or link delay even when signal strength would have been a more reliable predictor of throughput. This can result in node selections corresponding to lower throughput but improvements in delay cost and access-node load. However, as the number of stations increases, the behavior of the network stabilizes. The variability in RSSI, access-node load, and delay becomes more structured, and congestion patterns emerge. Moreover, more handover opportunities occur as user traffic increases, improving the chance of finding nodes that are both underutilized and have strong RSSI. In this context, Entropy’s adaptive weighting becomes increasingly effective, allowing it to respond to real-time network dynamics and avoid bottlenecks. Consequently, Entropy’s throughput performance improves and converges with that of the RSSI baseline at higher densities.

Overall, these results show that both AHP and Entropy are scalable, context-aware methods that adapt to network conditions. AHP consistently outperforms the traditional RSSI-based approach by making more intelligent handover decisions, while Entropy demonstrates strong adaptability under higher network loads. Their performance advantages reinforce the value of multi-criteria handover strategies in managing throughput in heterogeneous, multi-RAT environments.

\subsubsection{Summary} The results show that \chrome{} framework using both the weighing methods, AHP and entropy, outperforms the RSSI based handover in terms of Load on 5G network, Number of handovers, Delay cost. The AHP based \chrome{} framework outperforms the RSSI based handover in throughput and Entropy based \chrome{} framework slightly underperforms for cases of lower number of stations, but catches up in cases of higher traffic. Based on these observations, the following can be concluded:

\begin{itemize}
    \item \textbf{Efficient Offloading:} The combination of reduced 5G network load and handovers, especially handovers to 5G, highlights the efficiency of both AHP and Entropy in offloading traffic to alternate access technologies. This balanced redirection reduces peak congestion and supports more stable, scalable connectivity in multi-RAT environments.
    
    \item \textbf{QoS Improvement via Delay Reduction:} Both MCDM methods provide a substantial reduction in average link delay after handover, with modest improvements or slightly lower but sufficient throughput (in Entropy's case at low densities). This improvement in delay, when paired with effective offloading, demonstrates that the proposed algorithms not only relieve cellular network load but also enhance end-user Quality of Service during handover events.

\end{itemize}

The following can be concluded about the applicability of the AHP and entropy-based MCDM offloading algorithms:

\begin{itemize}
    
    \item \textbf{AHP based\chrome{} framework: Robust Performance in Unpredictable Conditions} \\
    AHP based\chrome{} framework is particularly effective in scenarios where network conditions are uncertain or volatile—for example, environments with high variability in RSSI or erratic delay patterns. Its fixed-priority structure (e.g., RSSI $>$ Load $>$ Delay) ensures that handover decisions remain anchored to signal strength, while still accounting for congestion and latency. This makes AHP highly reliable when sudden drops in RSSI could otherwise mislead more dynamic algorithms. As discussed in section ~\ref{sec_perfAnalysis_results}, the AHP based \chrome{} framework consistently outperformed the RSSI-based handover across all metrics. It reduced 5G load, minimized the number of handovers, and maintained a negative delay cost. Notably, in the 64-station scenario, the AHP based\chrome{} framework reduced WiFi to 5G handovers by 55\% and improved throughput by 11\% over RSSI, all while significantly lowering delay cost by 123.34\%. This shows that AHP based\chrome{} framework strikes an effective balance between stability and performance, even as network density increases. 

    \item \textbf{Entropy based \chrome{} framework: High Adaptability in Structured, High-Load Scenarios} \\
    Entropy based \chrome{} framework excels in environments where the network exhibits predictable patterns, such as stable RSSI and more structured variability in access-node load and link delay. Its dynamic weighting mechanism adjusts in real time to prioritize the most informative metric, allowing it to make context-aware handover decisions. This adaptability is especially advantageous in high-density, high-traffic scenarios where load balancing and delay mitigation are more critical than raw signal strength. Entropy based \chrome{} framework is particularly valuable in scenarios where throughput is not the sole objective, and delay-sensitive applications (e.g., video calls, VoIP, or gaming) require dynamic optimization of latency and congestion. It enables the network to shift focus toward real-time conditions, making it suitable for intelligent traffic steering in dense urban or enterprise deployments with known usage patterns.
    \item \textbf{MCDM-Guided Contextual Offloading:} As shown in Section~\ref{sec_perfAnalysis_usecase2}, in scenarios where the 5G base station exhibits high transmit power but is heavily loaded, the use of simple AHP and Entropy-based MCDM handover does not guarantee offloading. In such cases, the proposed \chrome{} framework, illustrated in Fig.~\ref{HO_mcdm_algo}, ensures that a WiFi AP with a high MCDM ranking—obtained through AHP or Entropy-based TOPSIS—is selected for offloading when it meets a minimum performance threshold. This approach helps avoid defaulting to overloaded 5G nodes due to RSSI dominance and instead redirects traffic to less congested alternatives. Importantly, this rank-based offloading strategy was not only validated through use-case analysis in section ~\ref{sec_perfAnalysis_usecase2} but also contributed directly to the performance improvements observed in the simulation-based evaluation. The inclusion of this mechanism enhanced the ability of the \chrome{} framework to reduce load on the 5G network, lower delay, and maintain throughput, reinforcing the effectiveness of the overall offloading framework.
    
\end{itemize}

\section{Conclusion and Future Works} \label{sec_conclusion}

This work tackled handover and offloading in heterogeneous multi-RAT networks by proposing \chrome{}, a generalized MCDM-based framework that unifies subjective (AHP) and objective (entropy) weighting within a TOPSIS pipeline and augments it with a simple, method-agnostic selection safeguard. Our analysis leads to three key findings. First, traditional RSSI based handover is insufficient in dense, heterogeneous deployments as higher-power macro links (e.g., 5G gNBs) tend to dominate RSSI and mask viable local-access options (e.g., WiFi APs), preventing congestion-relieving selections. Second, multi-criteria decision making alone does not guarantee congestion-aware handover or offloading. As shown in our use cases, entropy-based weights can overemphasize whichever metric is most dispersive (often RSSI), while AHP requires manual retuning (e.g., load-first) and can be unstable when RSSI is marginal. Both can select an overloaded node under congestion. Third, to address these limitations we propose \chrome{}, which uses MCDM as the ranking mechanism using RSSI, access-node load, link delay and adds a RAT-based RSSI threshold, to ensure that the alternative offloading target can maintain a feasible connection. We also add a rule that, when the top-ranked node is overloaded, traffic is redirected to the nest best qualifying AP to relieve congestion. This keeps the decision process policy-aligned (AHP) or data-adaptive (entropy) while enforcing a lightweight feasibility constraint that steers offloads away from congestion.

Evaluation of \chrome{} framework via Mininet-WiFi emulation across varying network densities show that both AHP-TOPSIS and Entropy-TOPSIS, when used within the \chrome{} framework, improved performance relative to RSSI-based handover. In the 64-station scenario, \chrome{} framework with AHP reduced 5G load by 31\%, WiFi to 5G handovers by 55\%, and delay cost by 123\%. \chrome{} framework with entropy achieved a 32\% reduction in 5G load, a 54\% reduction in handovers, and a 166\% improvement in delay cost. Throughput changes were modest, but the collective gains—lower macro cell load, fewer cross-RAT handovers, and consistently negative delay costs—demonstrate the advantage of a multi-metric, guard-protected approach over single-criterion selection.

Therefore, we show that (i) RSSI-only handovers are inadequate for heterogeneous, congestion-prone multi-RAT environments; (ii) unguarded MCDM based handovers can still favor overloaded links; and (iii) combining AHP/entropy weighting with a RAT-based RSSI threshold yields stable, congestion-aware offloading without method switching in runtime. Future work includes implementation of \chrome{} across other multi-RAT deployments, analysis with RATs other than 5G and WiFi such as vehicular networks, IoT, satellite networks, use of adaptive thresholds, extending to additional criteria (e.g., jitter, bandwidth), and reinforcement of guard logic to further reduce ping-pong effect while preserving responsiveness.

\section*{Acknowledgment}

This material is based upon work supported by the National Science Foundation under Grant Number 2030122.

\bibliography{references.bib}

\end{document}